\documentclass[preprint,12pt]{elsarticle}

 \usepackage{amssymb,amsmath}
\usepackage{graphicx}
\usepackage{epstopdf}
\usepackage{bm}

\begin{document}

\begin{frontmatter}



\title{Bose gas:  Theory and Experiment}


\author{Alexander L.\ Fetter}
\ead{fetter@stanford.edu}
\address{Geballe Laboratory for Advanced Materials\\
Departments of Physics and Applied Physics\\
Stanford University, Stanford, CA 94305-4045}

\author{Christopher J.\ Foot}

\address{Clarendon Laboratory\\
Parks Road\\
Oxford.\ OX1 3PU United Kingdom}
\begin{abstract}

For many years,  $^4$He typified Bose-Einstein superfluids, but recent  advances in dilute ultra-cold alkali-metal gases have provided  new   neutral superfluids that are particularly tractable because the system is dilute.  This chapter starts with a brief review of the physics of superfluid $^4$He, followed by the  basic ideas of Bose-Einstein condensation (BEC),  first for  an ideal Bose gas and then considering the effect of interparticle interactions, including time-dependent phenomena.  Extensions to more exotic condensates include magnetic dipolar gases, mixtures of two components, and spinor condensates that require a focused infrared laser for trapping of all the various hyperfine magnetic states in a particular  hyperfine $F$ manifold of $m_F$ states.   With an applied rotation, the trapped BECs nucleate quantized vortices.  Recent theory and experiment have shown that  laser coupling fields can mimic the effect of rotation.  The resulting synthetic gauge fields have produced vortices in a nonrotating condensate.  
\end{abstract}


\end{frontmatter}

\tableofcontents

\section{Introduction to Bose superfluids:  brief review of superfluid  $^4$He}
\label{intro}

Superfluid $^4$He has played a central role in understanding the physics of Bose-Einstein condensation.  It  has the unusual property of remaining liquid to $T = 0$ K under atmospheric pressure.  This behavior reflects its weak interatomic potential and the relatively large zero-point energy (because of the small mass).  The common helium isotope $^4$He has two electrons, two protons and two neutrons and thus acts like  a boson under exchange of two such atoms. Below a characteristic temperature $T_\lambda\approx  2.17$~K, it undergoes a phase transition from a normal  weakly viscous fluid to a superfluid with remarkable properties (for a review of this fascinating system, see~\cite{Till90}).

\subsection{two-fluid model}
\label{2fl}
Superfluid $^4$He has
remarkable hydrodynamic properties.  It can flow through fine channels with no pressure drop, which suggests that it has zero viscosity, yet a direct measurement of the viscosity through torque on  a rotating cylinder yields a value comparable with that for the normal phase above $T_\lambda$.  This seemingly contradictory behavior was explained phenomenologically by a  two-fluid model, in which the low-temperature phase has two interpenetrating components:  (1) a superfluid with zero viscosity and   irrotational flow velocity $\bm v_s$ with  $\bm\nabla \bm \times \bm v_s = 0$ and (2) a normal (viscous) fluid with general velocity $\bm v_n$.  Each component has its own temperature-dependent mass density $\rho_s$ and $\rho_n$, with $\rho_s +\rho_n =\rho$ (the total mass density).  The superfluid component has $\rho_s = 0$  at  $T_\lambda$ and $\rho_s = \rho$ at $T = 0$ K.  The normal density $\rho_n$ has been measured by the viscous drag on a set of oscillating disks that respond only to the normal component;   the superfluid density is then given by $\rho_s = \rho - \rho_n$.

Superfluid $^4$He can exhibit persistent currents in a multiply connected geometry like a torus or in a porous random medium such as a compacted   powder.  If the container and fluid rotate above $T_\lambda$ and are then cooled below $T_\lambda$ the superfluid will continue to rotate at essentially  the original angular velocity $\Omega$.  If the container is slowly brought to rest, in many cases, the superfluid continues to rotate  (in contrast to the behavior of a classical viscous fluid that would eventually come to rest).  Such persistent currents typify both superfluid $^4$He and superconductors (where the charge makes the detection of such electrical currents relatively easy).  There is a critical velocity $v_c$  for superfluid flow that is set by the energy of the lowest excited state;  $v_c$   gives an upper limit to the magnitude of persistent currents in confined geometries. For a channel of lateral dimension $\sim d$, the typical observed critical velocity is of order $v_c\sim \hbar/Md$, where $M$ is the atomic mass.

\subsection{quantized circulation}\label{circ}

For low flow velocity, liquid $^4$He is effectively incompressible with $\bm \nabla \cdot \bm v_s=0$.  Since $\bm v_s$ is also irrotational,  it  can be described with a velocity potential $\Phi$ that satisfies Laplace's equation, with $\bm v_s =\bm \nabla \Phi$.   For a quantum fluid, we can be more specific,  with  $\Phi =  (\hbar/M) S$, where $M$ is the atomic mass and $S$ is the phase of a macroscopic one-body wave function  characterizing the quantum-mechanical condensed state of superfluid $^4$He at $T=0$ K.  The irrotational condition on $\bm v_s$ apparently suggests that the superfluid would not rotate at $T = 0$ K, but experiments showed that the rotating superfluid had a parabolic meniscus independent of temperature (like that of a classical viscous fluid).  In 1949, Onsager proposed without explanation that the superfluid circulation  $\kappa =\oint d\bm l \cdot \bm v_s $ was quantized in units of $h/M$.  Subsequently  Feynman~\cite{Feyn55} suggested that rotating superfluid $^4$He has a uniform  array of quantized vortex lines with singular vorticity $\kappa$ at the center of each line.\footnote{Donnelly~\cite{Donn91}  gives  good account of this intriguing early history in Sec.~2.3.}  The velocity potential for a single vortex line is proportional to the azimuthal angle $\phi$, which ensures that the quantum-mechanical wave function is single-valued and reproduces Onsager's quantized circulation.  Feynman also chose the areal vortex density $n_v$ to mimic solid-body rotation $\bm v_{\rm sb} = \bm \Omega\times \bm r$, yielding $n_v= M\Omega/\pi \hbar$, where $\Omega $ is the external angular velocity.

\subsection{quasiparticles and the Landau critical velocity}\label{quasi}

Landau~\cite{Land41,Land47}  introduced a quasiparticle model to describe the normal component (for more accessible accounts, see~\cite{Till90,Lifs80}).  Specifically the elementary excitations act like phonons at long wavelengths with a dispersion relation
$\omega_k = sk$,
where $s\approx 240$ m/s is the speed of compressional sound.  For shorter wavelengths comparable with the interparticle spacing in liquid $^4$He, the dispersion relation has a local minimum associated with what are called rotons
\begin{equation}\label{roton}
\omega_k \approx\frac{ \Delta}{\hbar} + \hbar \frac{(k-k_0)^2}{2M^*},
\end{equation}
where $\Delta/k_B \approx 8.7 $ K is the roton gap, $k_0 \approx 1.9\times 10^{10} $ m$^{-1}$, and $M^* \approx 0.16 M$ for $^4{\rm He}$.  This model of independent phonons and rotons allows a straightforward calculation of the temperature-dependent normal-fluid density $\rho_n$.  More importantly, it also predicts a critical velocity as follows.  Imagine a macroscopic object moving through the superfluid at $T = 0$ K.  Conservation of energy and momentum indicates that this object cannot lose energy by creating a quasiparticle unless its speed exceeds a critical velocity $v_c$ given by the minimum value of $\omega_k/k$:
\begin{equation}\label{vc}
v_c = \left. \frac{\omega_k}{k}\right|_{\rm min}.
\end{equation}
  For superfluid $^4$He,  rotons determine this value as $v_c \approx \Delta/\hbar k_0\approx 50 $ m/s.

\section{Three-dimensional  ideal Bose gas}\label{3d}

This section provides a brief review of the ideal Bose gas, both the familiar uniform case and for a gas confined in a harmonic trap as in most experiments with ultracold atomic vapors (good general references are ~\cite{Dalf99,Pita03,Peth08}).

\subsection{qualitative picture of Bose-Einstein condensation in a uniform system}\label{Sec_qualpicBEC}

Consider a uniform ideal gas with number density $n$ and volume per particle $n^{-1}$.  One important characteristic length is the interparticle spacing $\sim n^{-1/3}$.  Suppose that the system is in thermal equilibrium at temperature $T$.  If $M$ is the particle mass,  the mean thermal momentum is $p_T\sim \sqrt{Mk_B T}$.  The de Broglie relation then gives the  thermal wavelength $\lambda_T\sim \hbar/\sqrt{Mk_B T}$, which provides another characteristic length for this uniform ideal gas.

In the classical limit $\hbar\to 0$ or at high temperature, the
thermal wavelength becomes small compared to the interparticle
spacing $n^{-1/3}$, which means that the effect of quantum
diffraction and interference is negligible.  This limit is
equivalent to ray optics for light when the wavelength is very small
compared to the typical dimension.

As $T$ falls, however, the thermal wavelength eventually becomes
comparable with $n^{-1/3}$, and quantum interference between nearby
particles becomes crucial.  It is convenient to introduce the
dimensionless  phase-space density $n\lambda_T^3$, which is
small in the classical limit.  When $n\lambda_T^3$ is of order
unity, however, the ideal gas changes character.  An ideal  gas of
fermions undergoes a crossover from the classical ideal gas to  a
degenerate Fermi gas at the Fermi temperature $T_F$  given by the
qualitative relation $n\lambda_{T_F}^3\sim 1$.  Equivalently,
$k_BT_F \sim \hbar^2n^{2/3}/M$, is the usual expression in
condensed-matter physics.  For $n\sim 10^{28} $ m$^{-3}$ and $M$
equal to the electron mass, the Fermi temperature $T_F$ for
electrons in a metal  is of order $10^4$ K, whereas liquid $^3$He
has similar number density but particle mass $\sim 10^4$ times
larger, yielding $T_F \sim 1$ K.

In contrast, an ideal gas of bosons in three dimensions undergoes a
{\em sharp} phase transition at a temperature $T_c$ given by the
similar criterion: $n\lambda_{T_c}^3\sim 1$, or equivalently  $k_B T_c\sim \hbar^2n^{2/3}/M$, similar to that for a
Fermi gas.  For liquid $^4$He, the number density and mass are
comparable to those for $^3$He, so that the transition temperature
is of order $1$ K. Nevertheless, the physics of the Bose-Einstein
condensation is quite different from the onset of Fermi degeneracy,
and we now explore the specific cases of a  Bose gas in a box and in
a general harmonic trap.
\subsection{quantitative description of ideal Bose gas in an external potential $V_{\rm tr}$}

Consider an external trap with potential $V_{\rm tr}$, and an associated complete set of quantum-mechanical single-particle states with energies $\epsilon_j$.  Assume an ideal Bose gas in equilibrium at temperature $T$ and chemical potential $\mu$ in this external potential.  The mean occupation number of the state $j$ is
\begin{equation}\label{occ}
n_j = \frac{1}{\exp[\beta(\epsilon_j-\mu)]-1} \equiv f\left(\epsilon_j\right),
\end{equation}
where $\beta^{-1} = k_B T$, and $f(\epsilon) = \{\exp[\beta(\epsilon-\mu)]-1\}^{-1}$ is the usual Bose-Einstein distribution function.  The mean total particle number and mean total energy are given by
\begin{equation}\label{N}
N(T,\mu) = \sum_j f(\epsilon_j),
\end{equation}
\begin{equation}\label{E}
E(T,\mu) = \sum_j \epsilon_j f(\epsilon_j).
\end{equation}
Formally, one can invert Eq.~(\ref{N}) to express the chemical potential as $\mu(T,N)$, thus giving  the energy in terms of the  more familiar  variables $T,N$.

Since we shall deal with many states, it is convenient to introduce the density of states $g(\epsilon) =\sum_j \delta(\epsilon -\epsilon_j)$ where the sum is over all distinct single-particle states (some of which may be degenerate). This applies for any particular trap potential.  Equation (\ref{N}) 
can then be rewritten as
\begin{equation}\label{Ng}
N(T,\mu) = \int d\epsilon\, g(\epsilon) \,f(\epsilon).
\end{equation}

In the classical limit for fixed $N$, the chemical potential is large and negative [since $n_j \ll 1$ in Eq.~(\ref{occ})], but as the temperature decreases, the chemical potential increases and eventually approaches the lowest single-particle energy $\epsilon_0$.  This equality $\mu(T_c,N)=\epsilon_0$  defines the critical temperature for the onset of Bose-Einstein condensation (BEC).  For $T<T_c$, the chemical potential cannot increase, because the distribution function $f(\epsilon_0)$ would become singular.  In contrast, Eq.~(\ref{Ng}) continues to decrease for $T<T_c$.  The explanation is that a macroscopic number of particles starts to occupy the lowest single-particle state with occupation number $N_0(T)$.  The temperature dependence of $N_0(T)$ follows from conservation of particles $N = N_0(T) + N'(T)$, where
\begin{equation}\label{N'}
N'(T) = \int_{\epsilon_0}^\infty d\epsilon\,\frac{g(\epsilon)}{\exp[\beta(\epsilon-\epsilon_0)]-1}
\end{equation}
defines the total number of particles not in the condensate.  Note that this integral is finite only if $g(\epsilon_0)$ vanishes; otherwise $N'(T)$ diverges and BEC cannot occur.

\subsection{ideal Bose gas in three-dimensional box with periodic boundary conditions}

Imagine a cubical box of length $L$ on a side.  Plane waves have the familiar normalized wave function $\psi(\bm r) = V^{-1/2} \exp(i\bm k\cdot \bm r)$ with $V = L^3$ and single-particle energy $\epsilon_{\bm k} = \hbar^2 k^2/2M$.  The states obey periodic boundary conditions for the specific wave vectors $\bm k = (2\pi/L)\,(n_x,n_y,n_z)$, where $n_j$ is any integer.  The lowest single-particle state is uniform with $\bm k = \bm 0$ and $\epsilon_0 = 0$.  The corresponding density of states is readily found to be
\begin{equation}\label{gbox}
g(\epsilon) = \frac{V}{4\pi^2}\,\left(\frac{2M}{\hbar^2}\right)^{3/2} \epsilon^{1/2},
\end{equation}
which vanishes at $\epsilon = 0$.  The condition for the onset of BEC yields the familiar condition for the phase-space density $n\lambda_{T_c}^3 = \zeta(\frac{3}{2}) \approx 2.612$ at $T_c$.

Below $T_c$, the fraction of particles in the excited states (those with finite momentum) is given by
\begin{equation}\label{N'box}
\frac{N'(T)}{N} = \left(\frac{T}{T_c}\right)^{3/2}.
\end{equation}
Conservation of total particles then yields the fraction of condensed particles in the  zero-momentum single-particle ground state
\begin{equation}\label{N0box}
\frac{N_0(T)}{N} = 1 - \left(\frac{T}{T_c}\right)^{3/2},
\end{equation}
which
increases continuously from 0 at $T_c$  to 1 at $T = 0$ K.

\subsection{ideal Bose gas in three-dimensional harmonic trap}

Typical experiments on dilute alkali-metal gases rely on magnetic traps that vary  quadratically  with the distance from the origin (see Chap.~1).   Hence we consider a general three-dimensional harmonic  trap  potential
\begin{equation}\label{Vharm}
V_{\rm tr}(\bm r) = \textstyle{\frac{1}{2}} M\left(\omega_x^2 x^2 + \omega_y^2 y^2 +\omega_z^2 z^2\right)
\end{equation}
with the familiar single-particle energies
\begin{equation}\label{epsharm}
\epsilon_{n_x n_y n_z} = \hbar\left(n_x\omega_x + n_y\omega_y + n_z\omega_z\right) + \epsilon_0,
\end{equation}
where $n_x, n_y, n_z$ are non-negative integers and $\epsilon_0 = \frac{1}{2}\hbar(\omega_x + \omega_y + \omega_z)$ is the zero-point energy in this harmonic trap.  The ground-state wave function $\psi_0(x,y,z)$ is a product of three Gaussians with characteristic dimensions $d_j = \sqrt{\hbar/M\omega_j}$, where $j = x, y, z$.  If the sums in the density of states are approximated by integrals (which holds for $\epsilon \gg \epsilon_0$), the resulting density of states becomes
\begin{equation}\label{gharm}
g(\epsilon) = \frac{\epsilon^2}{2\hbar^3 \omega_0^3},
\end{equation}
where $\omega_0 = (\omega_x\omega_y\omega_z)^{1/3}$ is the geometric-mean frequency.

The onset of  BEC in this harmonic trap occurs for $\mu(T_c,N) = \epsilon_0$, with the transition temperature
\begin{equation}\label{Tcharm}
k_BT_c = [\zeta(3)]^{-1/3}\hbar\omega_0 N^{1/3}\approx 0.94\, \hbar\omega_0 N^{1/3},
\end{equation}
where $\zeta(3) \approx 1.212$.  Below $T_c$, the number of
particles in the excited states decreases like $N'(T)/N= (T/T_c)^3$.
The remaining particles occupy the ground state so
\begin{equation}\label{N0harm} \frac{N_0(T)}{N}=
1-\left(\frac{T}{T_c}\right)^3.
\end{equation}
The condensate fraction grows from 0 at $T_c$ to 1 at $T=0$\,K as in
the case of a uniform Bose gas, but with a temperature dependence
different from that in Eq.\ (\ref{N'box}). The single-particle
ground-state wave function is that of a harmonic oscillator,
$\psi_0(x,y,z)\propto \prod_j\exp(-x_j^2/2d_j^2)$. Note that the
nontrivial trapping potential $V_{\rm tr}$ introduces a new
characteristic oscillator length $d_0 = \sqrt{\hbar/M\omega_0}$ that
does not appear for the uniform ideal Bose gas. For a noninteracting
gas (or small numbers of atoms) the zero-temperature condensate
density has a Gaussian profile $n_0(\bm r) =N |\psi_0(\bm r)|^2$.
For an interacting gas the size of the condensate is larger than the
harmonic oscillator length as shown in Sec.\
\ref{Sec_static_behavior}, but Eq.\ (\ref{N0harm}) remains a good
approximation since it is the Bose-Einstein  statistics that determine $T_c$,  not the interactions.

Images of the temperature-dependent condensate~\cite{Ande95}
provided clear evidence for the formation of a BEC in dilute cold
alkali-metal gases (see, for example,~\cite{Dalf99,Pita03,Peth08}).
Typical atomic traps have $d_0\sim $ a few $\mu$m, and $N\sim 10^6$,
leading the transition temperatures of order 100-1000 nK, depending
on the atomic mass.

\subsection{ideal Bose gas in  harmonic traps with reduced dimensions}

The shape of the confining potentials for ultracold atoms (created
by applied magnetic fields, laser light or other external fields)
can be engineered to be highly asymmetric so that certain degrees of
freedom are frozen out. Strong confinement in two directions,
$\omega_z \ll \omega_x \simeq \omega_y$, creates a long thin tube of
atoms and such one-dimensional systems have some fascinating
properties. For example in a Tonks-Giradeau gas (see Chap.~4)
bosonic atoms  act like fermions (in some but not all ways).
This behavior arises when the confinement is so tight that atoms
cannot pass each other, and  this one-dimensional system of
impenetrable bosons resembles beads on a string, or cars in a
traffic jam---atoms that cannot be at the same position along the
system look as if they  obey the Pauli exclusion principle.

Conversely for very tight confinement in one direction, $\omega_z
\gg \omega_x \simeq \omega_y$ in Eq.~(\ref{Vharm}), the cloud of
atoms has a oblate (pancake) shape; when the temperature is low
enough $k_{\mathrm B} T \ll \hbar\omega_z$ the atoms move only in
the radial plane  giving a two-dimensional gas, e.g.\ for $\omega_z
/2\pi = 2$\,kHz this requires $T \ll 100$\,nK which is readily
achievable. Two-dimensional behavior has been investigated in films
of superfluid helium but an important difference is that the helium
system corresponds to a square well (with a flat potential between
the walls) whereas in ultracold gases the atoms are usually in a
harmonic potential---these two situations have different densities
of the states [as we have seen for the three-dimensional cases in
Eq.~(\ref{gbox}) and Eq.~(\ref{gharm})]. A two-dimensional gas in a
uniform potential is a special case: the density of states is
independent of energy so the integral that gives $N'$ diverges in
Eq.~(\ref{N'})---this means that BEC does not occur in this case.
Nevertheless  interesting phenomena occur in a two-dimensional gas
in the quantum degenerate regime such as the
Berezinskii-Kosterlitz-Thouless transition, when the density of
atoms per unit area is sufficiently low, and the physics related to
this is discussed in Chap.\ 4.   Other trapping geometries
have also been demonstrated, one of the most interesting being the
creation of ring-shaped clouds where there can be persistent flow
around the loop, as for superfluid helium in similar multiply
connected systems. Another case is that of rapid rotation about the
axis
 of a cylindrically symmetrical potential which is described described in more detail in
Sec.~\ref{rot}; centripetal acceleration causes the gas cloud to bulge outward along its equator so that it can become flattened into the
2-D regime, i.e.\ effectively the tightness of trap is reduced in radial direction in the rotating frame, as shown in Eq.~(\ref{eqn:rot_hamilt}).

\section{Energies and length scales for interacting Bose gas}

The inclusion of a two-body interaction potential $U(\bm r - \bm
r')$ leads to several new features that play a crucial role in the
description of realistic dilute cold Bose gases.  Typically, the
interactions have a short range much less that the interparticle
spacing, and, at low temperature, only $s$-wave scattering is
important.  To incorporate these restrictions, it is convenient to
introduce a simple model, approximating the isotropic interparticle
potential as  $U(r) = g\delta^{(3)}(\bm r)$, where $g$ has the
dimension of energy times volume.   Standard scattering theory for
two particles with identical masses $M$ yields the identification
\begin{equation}\label{g}
g = \frac{4\pi \hbar^2 a}{M},
\end{equation}
where $a$ is the $s$-wave scattering length that relates the phase
of the scattered wave to that of incident wave.  For the commonly
used  bosonic alkali-metal atoms   ($^{23}$Na, $^{87}$Rb), the
scattering length is positive and  a few nm (a common trapped  state of $^7$Li is unusual in
having a large negative scattering length that leads to very
different physics; see Sec.~\ref{attr}). Typically, the
dimensionless parameter $na^3$ that characterizes the ``diluteness''
of the gas  is generally  small.  In special cases,  however, the
scattering length is purposely enhanced with an applied magnetic
field, as occurs in  a ``Fano-Feshbach resonance'' (see Chap.\ 1).
\subsection{interaction energy for a uniform gas}

In an interacting gas with $N$ particles, the $i$th  particle at $\bm r_i$ experiences an effective mean-field  potential \begin{equation}\label{intpot}
V(\bm r_i) = g\sum_{j\neq i} \delta^{(3)}(\bm r_i-\bm r_j) \approx g n(\bm r_i),
\end{equation}
where the sum is over all other particles, and the last expression omits small corrections of order $1/N$.  This effective potential is just the one-body Hartree interaction with all the other particles $V_H(\bm r) = gn(\bm r)$,  familiar from atomic and condensed-matter physics.

 It takes an energy $E(N+1) - E(N) = gn$ to add one additional condensate particle to a uniform interacting dilute  condensed Bose gas in a box. This quantity  is just the chemical potential $\mu = (\partial E/\partial N)_V$ so that here $\mu = gn$.  For a uniform dilute gas in a box of volume $V$, the thermodynamic definition of $\mu$ implies the total energy $E(V,N) = \frac{1}{2} gN^2/V$.  Furthermore, the pressure is $p = -(\partial E/\partial V)_N = \frac{1}{2} gn^2$.

\subsection{healing length for a uniform gas}

In general, the kinetic energy operator is ${\cal T} = -\hbar^2 \nabla^2/2M$.  Since $\nabla^2$ has the dimension of an inverse squared length, comparison of the kinetic energy ${\cal T}$ and the Hartree energy  $V_H= gn $ leads to a characteristic squared length
\begin{equation}\label{xi}
\xi^2 = \frac{\hbar^2}{2Mgn} = \frac{1}{8\pi na},
\end{equation}
where the second form makes use of Eq.~(\ref{g}).
If a uniform Bose gas is perturbed locally (by an impurity or a rigid external boundary, for example), $\xi$ is the length over which the gas heals  back to its equilibrium density $n$.  In the dilute limit $na^3\ll 1$, the healing length $\xi$  is large compared to the interparticle spacing $n^{-1/3}$, confirming the mean-field character of the Hartree picture of the interaction.

\section{Gross-Pitaevskii picture for a trapped Bose gas}

At zero temperature, a {\em uniform} interacting Bose gas in a box has two microscopic length scales:  the scattering length $a$ and the interparticle spacing $n^{-1/3}$ (or equivalently, the healing length $\xi$).   In addition,  an interacting  {\em trapped} Bose gas has a length associated with the oscillator length $d_0$, which leads to many interesting new effects.  Although the original papers of Gross~\cite{Gros61} and Pitaevskii~\cite{Pita61}  focused on a uniform gas, Baym and  Pethick~\cite{Baym96} extended the treatment to a harmonic trap soon after the creation of a BEC in 1995~\cite{Ande95}.

\subsection{static behavior}\label{Sec_static_behavior}

Gross and Pitaevskii independently started from the time-independent Schr{\"o}dinger equation for the condensate wave function in an ideal gas and added the nonlinear Hartree potential, leading to a nonlinear Schr{\"o}dinger equation.  With the addition of the trap potential, the Gross-Pitaevskii (GP) equation becomes
\begin{equation}\label{GPE}
\left(-\frac{\hbar^2\nabla^2}{2M} + V_{\rm tr} + g|\Psi|^2\right)\Psi = \mu \Psi,
\end{equation}
where $\mu$ is the chemical potential and $\Psi$ is the condensate wave function.   Strictly, $\Psi$  is normalized  to the condensate number $N_0$, but  a dilute Bose gas at zero temperature has $N_0\approx N$ so that the condensate normalization is generally taken as $\int dV\,|\Psi|^2 \approx N$.

For a deeper understanding of this time-independent GP equation, it is helpful to start from the corresponding GP energy functional
\begin{equation}\label{EGP}
E_{\rm GP}[\Psi] = \int dV\left(\underbrace{\frac{\hbar^2|\bm \nabla \Psi|^2}{2M}}_{\rm kinetic} +\underbrace{ V_{\rm tr}|\Psi|^2}_{\rm trap}+\underbrace{\frac{1}{2}g|\Psi|^4}_{\rm interaction}\right)
\end{equation}
containing the kinetic energy, the trap energy (proportional to the condensate density), and the interaction energy (proportional to the condensate density squared).  Variation of $E_{\rm GP}[\Psi]-\mu N$ with respect to $\Psi^*$ at fixed normalization with the chemical potential as a Lagrange multiplier  reproduces the time-independent GP equation, Eq.~(\ref{GPE}); but note that although this looks like a nonlinear eigenvalue problem, the right-hand side of Eq.~(\ref{GPE}) contains the Lagrange multiplier $\mu$, rather than the energy per particle.

For a harmonic potential as in Eq.~(\ref{Vharm}), it is convenient to use the mean oscillator length $d_0$ as the length scale and   the mean oscillator frequency $\omega_0$ times $\hbar$ as the energy scale.  The resulting dimensionless form of the energy functional in Eq.~(\ref{EGP}) has a simple structure, with   the first two terms  of order unity (since they involve only the harmonic oscillator),  but the interaction term contains a new dimensionless parameter $Na/d_0$ that characterizes the importance of the interaction energy relative to the other two terms.  If this interaction parameter is small, then the BEC adjusts to balance the kinetic energy and the confining trap,   like an ideal gas, but the situation is very different if this parameter is large.

Note that the ratio $a/d_0$ is typically $\sim 10^{-3}$, but the total number of atoms is of order $10^5-10^6$, so that this interaction parameter is generally large for most experiments.   In this case, the repulsive interactions act to expand the condensate to a typical mean dimension $R_0$ that is considerably  larger than $d_0$ (typically $R_0/d_0\sim 10$ for $Na/d_0\sim 10^3$). Since this expanded condensate has a small density gradient, the first term of Eq.~(\ref{EGP}) becomes small, and the remaining terms give the approximate  Thomas-Fermi (TF) energy functional~\cite{Baym96}
\begin{equation}\label{ETF}
E_{\rm TF}[\Psi] = \int dV\left( V_{\rm tr}|\Psi|^2+\frac{1}{2}g|\Psi|^4\right),
\end{equation}
which involves only $|\Psi|^2$ and $|\Psi|^4$.  Variation of this  TF  energy functional with respect to $|\Psi|^2$ immediately yields the TF approximation
\begin{equation}\label{TF}
 V_{\rm tr}(\bm r) + g |\Psi(\bm r) |^2 = \mu,
\end{equation}
which also follows by omitting the kinetic energy in Eq.~(\ref{GPE}).  Solution of this algebraic equation gives the celebrated TF condensate density profile
\begin{equation}\label{nTF}
n(\bm r) = \frac{\mu - V_{\rm tr}(\bm r)}{g}\,\theta[\mu - V_{\rm tr}(\bm r)],
\end{equation}
where $\theta(x)$ is the unit positive step function.  For the general three-dimensional harmonic trap in Eq.~(\ref{Vharm}), this TF density is an inverted parabola
\begin{equation}\label{nharm}
n(\bm r) = n(0)\left(1-\frac{x^2}{R_x^2} -\frac{y^2}{R_y^2} -\frac{z^2}{R_z^2}\right)
\end{equation}
where the right-hand side is positive and zero otherwise.  Here, $n(0) = \mu/g$ is  the central density, and $R_j^2=2
\mu/M\omega_j^2$ are the squared condensate radii in the three orthogonal directions.

The normalization condition on the density $\int dV\, n(\bm r)\approx N$ yields $N=8\pi n(0)R_0^3/15$, where $R_0^3 = R_xR_yR_z$ depends on the chemical potential $\mu$~\cite{Baym96}.  Some analysis yields the important dimensionless relation
\begin{equation}\label{R0}
\frac{R_0^5}{d_0^5} = 15\frac{Na}{d_0},
\end{equation}
which is large in the present TF limit.  Similarly, the TF chemical potential becomes
\begin{equation}\label{muTF}
\mu_{\rm TF} = \frac{1}{2}M\omega_0^2 R_0^2 = \frac{1}{2} \hbar\omega_0\frac{R_0^2}{d_0^2},
\end{equation}
showing that $\mu_{\rm TF} \gg \hbar\omega_0$ in the TF limit.  Note that $\mu_{\rm TF} $ is proportional to $N^{2/5}$;  hence the thermodynamic relation $\mu = \partial E/\partial N$ yields the TF energy for the trapped condensate $E_{\rm TF} = \frac{5}{7} \mu_{\rm TF} N$.

It is natural to use the central density  $n(0)$ to define the healing length in a nonuniform condensate, and Eq.~(\ref{xi}) shows that $\xi^2=d_0^4/R_0^2$; equivalently,
\begin{equation}\label{geom}
\frac{\xi}{d_0}=\frac{d_0}{R_0}.
\end{equation}
The right-hand side is small in the TF limit, so that $\xi$ is also small compared to the mean oscillator length $d_0$.  This clear separation of TF length scales $\xi\ll d_0\ll R_0$ is frequently valuable in understanding the physics of the trapped TF condensate.

\subsection{time-dependent GP equation}

Equation (\ref{GPE}) has the intuitive time-dependent generalization (known as the time-dependent GP equation)
\begin{equation}\label{TDGP}
i\hbar\frac{\partial \Psi(\bm r,t)}{\partial t} = \left[-\frac{\hbar^2\nabla^2}{2M} + V_{\rm tr}(\bm r) + g|\Psi(\bm r,t)|^2\right]\Psi(\bm r,t),
\end{equation}
where $\Psi$ now depends on $t$ as well as on $\bm r$.  Comparison with Eq.~(\ref{GPE}) shows that a stationary solution has the time dependence $\exp(-i\mu t/\hbar)$.  To understand this time-dependence, note that the condensate wave function $ \Psi(\bm r,t) $ is a matrix element of the Heisenberg field operator $\psi(\bm r,t) = e^{iHt/\hbar}\psi(\bm r,0)e^{-iHt/\hbar}$. Since the operator $\psi$ removes one particle, the states on the right and left have $N$ and $N-1$ particles, respectively.  Their energy difference $E_N-E_{N-1}$ is  the chemical potential $\mu =\partial E/\partial N$.

This nonlinear field equation can be recast in an intuitive hydrodynamic form by writing the condensate wave function as
\begin{equation}\label{hyd}
\Psi(\bm r,t) = |\Psi(\bm r,t)|\exp[iS(\bm r,t)]
\end{equation}
in terms of the magnitude $|\Psi|$ and the phase $S$.  The condensate (particle) density is simply $n(\bm r,t) = |\Psi(\bm r,t)|^2$, and the usual one-body definition of the particle current density for the Schr{\"o}dinger equation shows that $\bm j = |\Psi|^2 \hbar\bm \nabla S/M$.  The hydrodynamic definition $\bm j = n\bm v$ then gives the local superfluid velocity as
\begin{equation}\label{vs}
\bm v(\bm r,t) = \frac{\hbar}{M}\bm \nabla S(\bm r,t) = \bm \nabla \Phi(\bm r,t),
\end{equation}
where $\Phi = \hbar S/M$ is the velocity potential for this irrotational flow.  The relation between the velocity $\bm v$ and the phase
$S$ implies that the circulation $\kappa = \oint_{\cal C} d\bm l\cdot \bm v$ around any closed path $\cal C$ in the fluid is quantized in units of $h/M$~\cite{Feyn55}
\begin{equation}\label{circ1}
\kappa = \frac{\hbar}{M} \oint_{\cal C} d\bm l\cdot \bm \nabla S= \frac{\hbar}{M} \Delta S_{\cal C},
\end{equation}
 because $\Delta S_{\cal C}$ must be an integral multiple of $2\pi$ to ensure that the condensate wave function $\Psi$ is single valued.

Substitute Eq.~(\ref{hyd}) into the time-dependent GP equation (\ref{TDGP}).  The imaginary part gives the expected conservation of particles
\begin{equation}\label{cons}
\frac{\partial n}{\partial t}+ \bm\nabla\cdot\left(n\bm v\right) = 0.
\end{equation}
In contrast the real part gives a generalized Bernoulli equation
\begin{equation}\label{Bern}
\frac{1}{2} M v^2 + V_{\rm tr} -\frac{\hbar^2}{2M\sqrt n}\nabla^2 \sqrt n + gn + M\frac{\partial \Phi}{\partial t} = 0.
\end{equation}
Here, the explicitly quantum-mechanical term involving $\hbar^2$ is called the quantum pressure.  It is usually small for low-lying collective modes of a trapped condensate in the TF limit, since the density is slowly varying.   It can be important for finite-wavelength disturbances, however,  as shown below for the Bogoliubov excitation spectrum (see, for example, Sec.~7.2 of~\cite{Peth08}).

\subsection{Bogoliubov spectrum:  linearized hydrodynamic equations for a uniform Bose gas}\label{hydroequations}

As a first application of these hydrodynamic equations, consider a uniform stationary Bose gas in a box with periodic boundary conditions (so that  $V_{\rm tr}$ acts solely to fix the allowed plane waves, but otherwise plays no role).
Assume small perturbations around the equilibrium number density $n_0$, writing $n\approx n_0 + n'$, where $n'$ is small.  Similarly the velocity potential becomes $\Phi_0 + \Phi'$.  The linearized  equation of continuity has the simple form
\begin{equation}\label{conslin}
\frac{\partial n'}{\partial t} + n_0 \nabla^2 \Phi' = 0
\end{equation}
because $\bm v' = \bm\nabla \Phi'$.

    To  zero order,  the Bernoulli equation yields  $gn_0 + M\partial \Phi_0/\partial t =0 $. Since $\Phi_0 = \hbar S_0/M$ and  $gn_0$ is simply the equilibrium chemical potential $\mu_0$ for a uniform Bose gas, this result reproduces the time-dependent phase of the equilibrium condensate wave function $\Psi \propto \exp(-i\mu_0 t/\hbar)$, as seen by comparing Eqs~(\ref{GPE}) and (\ref{TDGP}).

 In this example  all wavenumbers are relevant, and it is essential to retain the quantum pressure in the Bernoulli equation.  Thus, it is necessary to write $\sqrt{n} \approx \sqrt{n_0} + \frac{1}{2} n'/\sqrt{n_0}$.  The first-order contribution yields the additional linear equation
\begin{equation}\label{firstorder}
-\frac{\hbar^2 \nabla^2 n'}{4Mn_0} + gn' +M\frac{\partial \Phi'}{\partial t} = 0.
\end{equation}
Assume plane-wave solutions $\propto \exp i(\bm k\cdot\bm r - \omega t)$. The two linearized equations become
\begin{eqnarray}\label{lin}
-i\omega n' -n_0 k^2 \Phi' & = &0, \\[.2cm]
\left(\frac{\hbar^2 k^2}{4Mn_0} + g\right) n' -i\omega M \Phi'&=&0.
\end{eqnarray}
Solutions exist only if the determinant of coefficients vanishes, which leads to the celebrated Bogoliubov spectrum $\omega_k$ for a dilute Bose gas~\cite{Bogo47} (here written in terms of the excitation energy $\epsilon_k = \hbar\omega_k$)
\begin{equation}\label{Bog}
\epsilon_k^2 = \frac{ gn_0\hbar^2k^2}{M}  + \left(\frac{\hbar^2 k^2}{2M}\right)^2 =  \frac{ gn_0  \hbar^2k^2}{M} + \left(\epsilon_k^0\right)^2,
\end{equation}
where $\epsilon_k^0 = \hbar^2 k^2/2M$ is the free-particle excitation spectrum.  For a summary of Bogoliubov's original derivation based on quantized field amplitudes, see Sec.~7.2 of~\cite{Peth08}.

This dispersion relation $\omega_k = \epsilon_k/\hbar $  has two distinct regions depending on the wave number $k$.  At long wavelengths (small $k$), the spectrum is {\em linear} with $\epsilon_k \approx \hbar s k$, where $s = \sqrt{gn_0/M}$ is the speed of sound.  Note that $g$ must be positive for this uniform Bose gas, which means that the effective interparticle interaction must be repulsive.  In contrast, at short wavelengths (large $k$), the spectrum is quadratic, with $\epsilon_k \approx \epsilon_k^0 + gn_0$, shifted up from the free-particle value by the chemical potential $gn_0$ that reflects the presence of the background medium.

\begin{figure}[htbp]
\begin{center}
\includegraphics[scale = 0.8]{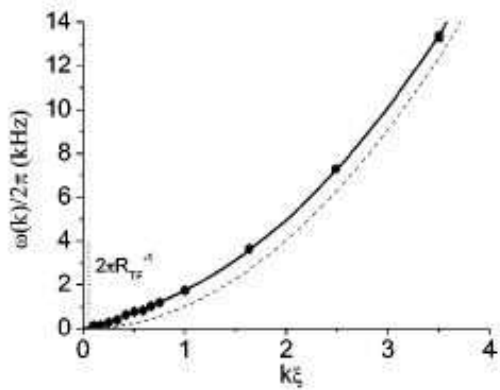}
\end{center}
\begin{flushleft}
\caption{\label{fig:Bog} Filled circles show measured excitation spectrum for dilute trapped BEC, with solid curve the Bogoliubov spectrum.  For comparison dashed curve shows the  free-particle quadratic spectrum.   Reprinted with permission of the authors~\cite{Ozer05} and the American Physical Society.} \end{flushleft}\end{figure}

The crossover between these two regimes occurs at $k_{\rm cr}^2  \approx 2Mn_0 g/\hbar ^2=1/\xi^2$, where $\xi= \hbar/\sqrt{2Mn_0g}$ is the healing length for the uniform gas [see Eq.~(\ref{xi})].  If there were  no interactions ($g\to 0$), then $k_{\rm cr}\to 0$, and the spectrum reduces to that of a free particle for all $k$.  Figure~\ref{fig:Bog} shows the measured Bogoliubov excitation spectrum  with no free parameters (the free-particle spectrum is the dashed line)

It is interesting to consider the Landau critical velocity (\ref{vc}) for this dilute interacting Bose gas.  Since the Bogoliubov spectrum is linear at small $k$ and changes to quadratic at large $k$, its derivative $d\omega_k/dk$ (the group velocity) is non-negative.  Hence the minimum of $\epsilon_k/\hbar k$ occurs at long wavelengths (in contrast to the case for superfluid $^4$He) and is simply the speed of sound
\begin{equation}\label{vcBog}
v_c \approx s = \sqrt{\frac{n_0g}{M}}.
\end{equation}
Note that  the repulsive interactions ($g>0$) are  essential to give a nonzero critical velocity.  For an ideal gas, the spectrum is quadratic for all $k$ and the corresponding critical velocity vanishes.  Thus an ideal Bose gas at zero temperature indeed has a Bose-Einstein condensate, but it is {\em not} a superfluid in the conventional sense because its critical velocity vanishes.

\subsection{linearized hydrodynamic equations for collective modes of a stationary condensate}

A different and very important application of the hydrodynamic equations is to study the small low-lying oscillations of an initially static condensate with density $n_0(\bm r)$.  In this case, the equation of continuity (\ref{cons}) becomes $\partial n'/\partial t + \bm\nabla\cdot (n_0  \bm v') = 0$, where $n'$ and $\bm v'$ are small perturbations.  Similarly, the linearized form of the Bernoulli equation (\ref{Bern}) yields $g n '+ M\partial  \Phi'/\partial t = 0$, where $\bm\nabla  \Phi'= \bm v' $.

The time derivative of the first of these equations and the gradient of the second readily yield the generalized wave equation
\begin{equation}\label{oscillations}
\frac{\partial^2 n'}{\partial t^2} =\bm \nabla \cdot \left(\frac{n_0(\bm r)g}{M}\bm \nabla  n'\right),
\end{equation}
which explicitly incorporates the equilibrium condensate density $n_0(\bm r)$ as an inhomogeneous effective potential.  Note that the quantity $gn_0(\bm r)/M$ can be identified as the local squared speed of sound $s^2(\bm r)$.  Hence  Eq.~(\ref{oscillations}) can equivalently be written
\begin{equation}\label{coll}
\frac{\partial^2 n'}{\partial t^2} =\bm \nabla \cdot \left[s^2(\bm r) \bm \nabla  n'\right],
\end{equation}
which is important in experimental studies of collective modes of trapped condensates, as described in Sec.~\ref{Sec_collective_modes}.

\subsection{effect of attractive interactions}\label{attr}

For liquid $^4$He, the interatomic potential is overall repulsive because of the strong repulsive core and the weak attractive van der Waals attraction. This is why it remains a liquid down to zero temperature and can only be solidified by going to high pressure (many  atmospheres) at low temperature.

The situation is very different for the alkali metals like Li, Na, K, and Rb, because each atom is much more polarizable than He.  As a result, the interatomic potential typically has many bound states, and the sign of the $s$-wave scattering length depends critically on the relative position of the last bound state~\cite{Peth08}.  For example, $^7$Li atoms have a  negative scattering length $a\approx  -1.46$ nm equivalent to an attractive interaction.

As noted in connection with Eq.~(\ref{vcBog}), a uniform dilute Bose gas with attractive interactions has an imaginary speed of sound and is thus unstable.  With a  trapped condensate, however,   the kinetic trapping energy already provides a positive energy that can counteract the negative attractive interaction energy.  For a quantitative  analysis, consider a spherical condensate with oscillator length $d_0$ and use a trial Gaussian wave function
\begin{equation}\label{trial}
\Psi(r) \propto \exp\left(-\frac{r^2}{2\beta^2 d_0^2}\right),
\end{equation}
where $\beta$ is a variational parameter (note that the condensate radius is $\beta d_0$).  The ground-state energy (\ref{EGP}) is easily evaluated to give
\begin{equation}\label{E_g}
E_g(\beta)  =\frac{1}{2} N \hbar \omega_0\left[\frac{3}{2}\left(\frac{1}{\beta^2} + \beta^2\right) +\sqrt{\frac{2}{\pi}}\,\frac{Na}{d_0}\frac{1}{\beta^3}\right],
\end{equation}
and $\beta < 1$ ($\beta>1$)  for attractive (repulsive) interactions.  Here, the three terms represent the kinetic energy, the trap energy, and the interaction energy, respectively. This treatment also gives a quantitative justification of the approximation made in going from Eq.~(\ref{EGP}) to (\ref{ETF}).

It is clear from inspection that $E_g(\beta)$ becomes large for $\beta \to \infty$ because of the spatial confinement of the harmonic trap.  The detailed behavior for  small $\beta$ depends crucially  on the sign and magnitude  of the interaction parameter $Na/d_0$.  For a  noninteracting gas with  $a = 0$, the minimum of Eq.~(\ref{E_g}) occurs for $\beta = 1$, which is just the familiar result for an oscillator. Baym and Pethick~\cite{Baym96} used this variational approach to study the expansion of the condensate  for strong repulsive interactions.  The situation is very different for an attractive interaction, no matter how weak, because the (now negative)  cubic term dominates for $\beta\to 0$, and $E_g(\beta)$ diverges to $-\infty$.  Hence  any attractive system is  globally unstable with respect to a collapse.  For small and moderate values of $N|a|/d_0$, however, $E_g(\beta)$  does retain a local minimum, and the system  then remains locally stable.  A  straightforward analysis shows that the minimum disappears at the critical value $N_c|a|/d_0 \approx 0.671$ and that the critical condensate radius is reduced by a factor $\beta_c = 5^{-1/4} \approx 0.669$.  For comparison, a numerical study of the GP equation~\cite{Rupr95} yields the value $N_c|a|/d_0 \approx 0.575$, which differs from the variational estimate by  $\approx 17\%$.

The attractive interaction in $^7$Li produces just this behavior.  For a trap size of order 3 $\mu$m, the critical stability limit is $N_c\sim 1250$ atoms~\cite{Sack99}.  In the experiments, atoms from the surrounding gas continue to augment the condensate until it reaches the critical value and then collapses.  This process repeats in a cyclic fashion.

\section{Selected applications and comparison with experiments}

The GP formalism, either in the form of the time-dependent equation (\ref{TDGP}) or the pair of coupled hydrodynamic equations (\ref{cons}) and (\ref{Bern}),  is remarkably rich and has many applications.  Here we consider a few of the most important ones.

\subsection{free expansion for different trap aspect ratios}

\begin{figure}[ht]
\begin{center}
\includegraphics[scale = 0.35]{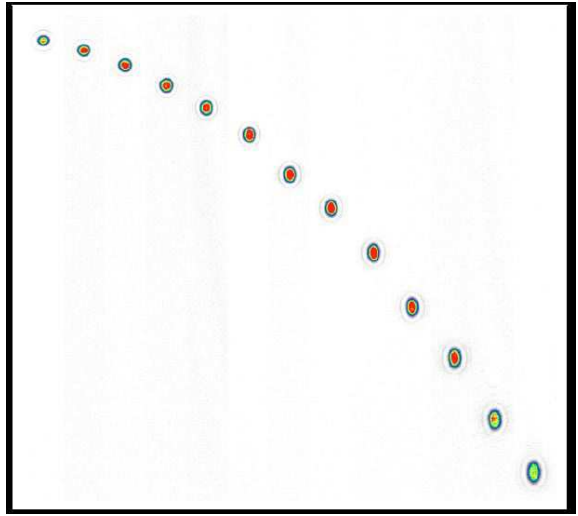}
\end{center}
\begin{flushleft}
\caption{\label{fig:BECexpansion} A montage of images showing how a
Bose-Einstein condensate expands after it has been released from a
trap and falls downwards under gravity.
The initial
elliptical cloud is elongated along the horizontal axis and is too
small to resolve properly. As it expands it becomes circular
about 7\,ms) 
and then elongated vertically. The final shape reflects
the initial momentum distribution of the BEC wave function and the
effects of interactions.   
Data provided by M. Gildemeister and B.
Sherlock, Oxford.} \end{flushleft}\end{figure}

Much of the experimental data relating to Bose-Einstein condensation in trapped gases
has been obtained from images of the density distribution of an atomic cloud after the
trap has been switched off. Indeed such time-of-flight expansion of an ultracold gas released
from a harmonic trap provided some of the first evidence of
Bose Einstein condensation \cite{Ande95}. In a series of observations taken a fixed time after release the measured size of the expanded cloud decreases abruptly as the
temperature is reduced through the phase transition.
A second characteristic sign of the transition is observed by recording the aspect ratio of
the atom cloud (the ratio $R_z/R_\perp$ of axial and radial widths) as it expands from an asymmetric trap; for a sufficiently long period of free expansion  (where the cloud in large compared to its initial size) a thermal cloud has an aspect ratio of unity (circular shape) while a BEC has an elliptical distribution (as shown in Fig.~\ref{fig:BECexpansion}). In the absence of interactions one can explain the difference above as follows.
The asymptotic density profile of an ideal gas released from a
potential reflects the initial momentum distribution of the cloud in the trap (when the distance $pt/M$ travelled by the atoms of momentum $p$ and mass $M$ in time $t$ is much greater than the initial size of the cloud).
A cloud of atoms that is far from the quantum regime has the same velocity distribution (Maxwell-Boltzmann) in all directions
characterized by the temperature, and hence expands isotropically.
On the other hand, atoms in a BEC occupy one quantum state and are described by a single wave function.
The momentum distribution is the Fourier transform of
the spatial wave function and has an aspect ratio of $\sqrt{\omega_{x}/\omega_{z}}$
when viewed along the $y$-direction. The anisotropy of the confining potential is thus reflected in the
aspect ratio of the expanding  BEC,  with the aspect ratio being \emph{inverted} with respect to its initial value in trap, i.e.\ the cloud expands most rapidly in
along the axis where the initial confinement is tightest, as expected from the Heisenberg uncertainty principle.
However interactions must be taken into account for a complete description of the expansion of a condensate.
They give rise to a mean-field interaction energy which is converted to
kinetic energy of motion during the release. For larger atom numbers this
interaction term, proportional to the density, dominates the
expansion of the condensate (as explained in Sec.~\ref{Sec_static_behavior}). Gradients in the density drive the initial acceleration of the atoms so the cloud expands
more rapidly in the most tightly confined direction. This causes the aspect ratio of the
cloud to invert during expansion as in the non-interacting case, but when interactions are dominant the calculation requires solving the set of
hydrodynamic equations given in Eq.~(\ref{cons}) and (\ref{Bern}),  with the external potential $V_{\rm{tr}}$ set to zero at the
release time (see also \cite{Baym96}). 
This approach based on mean-field theory predicted asymptotic aspect ratios for Bose Einstein condensates
that agree with the experimental results to within a few percent, and this  was one of the
first validations of the mean-field description of interactions in dilute atomic vapors.
\subsection{measurement of the condensate fraction}\label{Sec_condensate}

Superfluid $^4$He generally forms a macroscopic sample, which facilitates the study of the superfluid properties like the normal and superfluid densities and  persistent currents.  In contrast, it has proved extremely difficult to measure reliably the low-temperature  condensate fraction for bulk uniform superfluid $^4$He.  Conceptually, the true many-body ground-state wave function $\Psi_0$ can  be expanded in the complete set of states $|\Phi_i\rangle$ for an ideal Bose gas in a box.  The strong repulsive interactions for realistic superfluid $^4$He lead to a linear combination of many ideal-gas states with high energies in addition to the noninteracting ground state $\Phi_0$ with all the particles in the zero-momentum condensate.  Correspondingly,   the fraction of $^4$He atoms in the zero-momentum condensed state (the overlap $|\langle \Phi_0|\Psi_0\rangle|^2$) is significantly reduced, leading to an estimated condensate fraction $N_0/N\sim 0.1$, in contrast to the value 1 for an ideal uniform Bose gas.

Measurements of $N_0/N$ for superfluid $^4$He have relied on
quasi-elastic    scattering of high-energy neutrons.  In principle,
the single-particle momentum distribution function of bulk
superfluid liquid  helium has a delta function at $\bm k = \bm 0$,
with a weight proportional to $N_0/N$. Nevertheless,  the
experiments have proved challenging with considerable uncertainty.
Sokol~\cite{Soko95}  summarizes the experimental conclusions,
confirming the theoretical estimates of $N_0/N\sim 0.1$, and
Sec.~8.5 of Ref.~\cite{Pita03} gives an abbreviated account. The
situation is very different for a dilute ultracold gas.  Detection
of the  condensate fraction is reasonably straightforward and an
obvious characteristic feature of the onset  BEC in a trap is the
sudden appearance of the narrow condensate rising out of the broad
thermal cloud as $T$ falls below the actual transition
temperature~\cite{Ande95}. (Convincing direct measurements of the
superfluid properties of quantum degenerate gases followed a few
years after the first observation of BEC.)

\begin{figure}[h]
\includegraphics[width=3.0in]{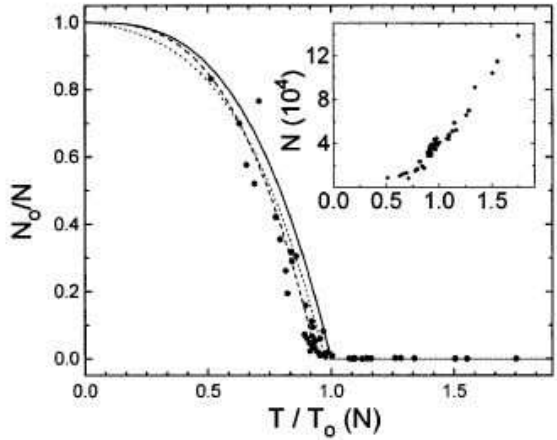}
\caption{ \label{condfrac} Scaled condensate-fraction ratio
$N_0(T)/N$ compared with that for an ideal gas in a general harmonic
trap.  Solid line is from Eq.~(\ref{N0harm}) with $T_0(N)$  the
transition temperature given in Eq.~(\ref{Tcharm}).  Measured data
fit a similar curve but with a reduced transition temperature
$0.94\, T_0(N)$ (dashed curve).  The inset shows the reduction in
the number of atoms $N$ as the system  cools. (Note that $N$ was
relatively small in this early experiment.) Reprinted with
permission of the authors~\cite{Ensh96} and the American Physical
Society.}
\end{figure} In the original experiment by the JILA
group~\cite{Ensh96}, and in many subsequent experiments, the atomic
cloud was allowed to expand for a definite time and then imaged by
recording the shadow of the cloud arising from absorption of a pulse
of resonant laser light. In general, the projected image has both
the broad  thermal background and the narrow condensate (for
$T<T_c$).  The ratio of the areas under these curves yields the
condensate fraction $N_0(T)/N$. Figure \ref{condfrac} shows the
measured values which extend down to $\sim 0.5\, T_c$.  Since the
gas is very dilute the condensate fraction closely resembles that
for an ideal trapped gas. Corrections from finite $N$ and many-body
interactions are estimated to be at most a few percent. In recent experiments~\cite{Tammuz2011} the condensation fraction has been measured in a system where the scattering length can be tuned to be close to zero so approximating very closely to an ideal gas; this improves the ability to observe condensates with small number of atoms and clarifies the interpretation of data like that shown in Fig.\ \ref{condfrac}. 

\subsection{low-lying collective oscillation modes}\label{Sec_collective_modes}

Ever since the first observation of BEC in a dilute gas the study of
collective oscillation modes has been crucial to our understanding of
these mesoscopic systems. For typical experimental conditions these excitations
dominate the low-frequency response of a weakly-interacting
gas to an external perturbation. Initial experiments performed at temperatures
well below $T_c$ tested the mean field theory based on the Gross Pitaevskii equation. The
very good agreement between theory and experiment with no free fit parameters provided strong evidence for the validity of this model. 
To model the system we return to the linearized hydrodynamic equation Eq.~(\ref{coll}).  
For a uniform system the solutions to Eq.\ (\ref{coll}) correspond
to sound waves propagating with the speed given in
Eq.~(\ref{vcBog}). In a trapped system the wavelength of the phonons
must be considered in relation to the dimensions of the cloud. For
wavelengths significantly less than all spatial dimensions of the
condensate the solutions are sound waves propagating at the local
sound velocity. Whereas for wavelengths comparable to the size of
the condensate, the boundary conditions produce standing sound waves
at discrete frequencies which are referred to as the collective
oscillation modes of the system. Thus the collective oscillation
frequencies of a BEC in a harmonic trap do not depend on
interactions (number of atoms) even though of the speed of sound $s$
does, i.e.\ $\omega=sk$ where the wavenumber $k$  is the same for
different numbers of atoms in the same trap (in the TF regime). This
arises because the speed of sound is $s \approx (n_0 g/M)^{1/2}$
(see Eq.~\ref{oscillations}), and the wavelength of the lowest
energy excitation is limited by the extent of the cloud in a given
direction $R_i$ [determined by $\mu = n_0 g = M\omega_i^2 R_i^2/2$
from Eq.~(\ref{muTF}) in the Thomas-Fermi regime]. Hence the lowest
mode has a frequency of $s/R_i \approx \omega_i$, which is the same
as for the non-interacting case.

\begin{figure}[htbp]
\begin{center}
\includegraphics[scale = 0.8]{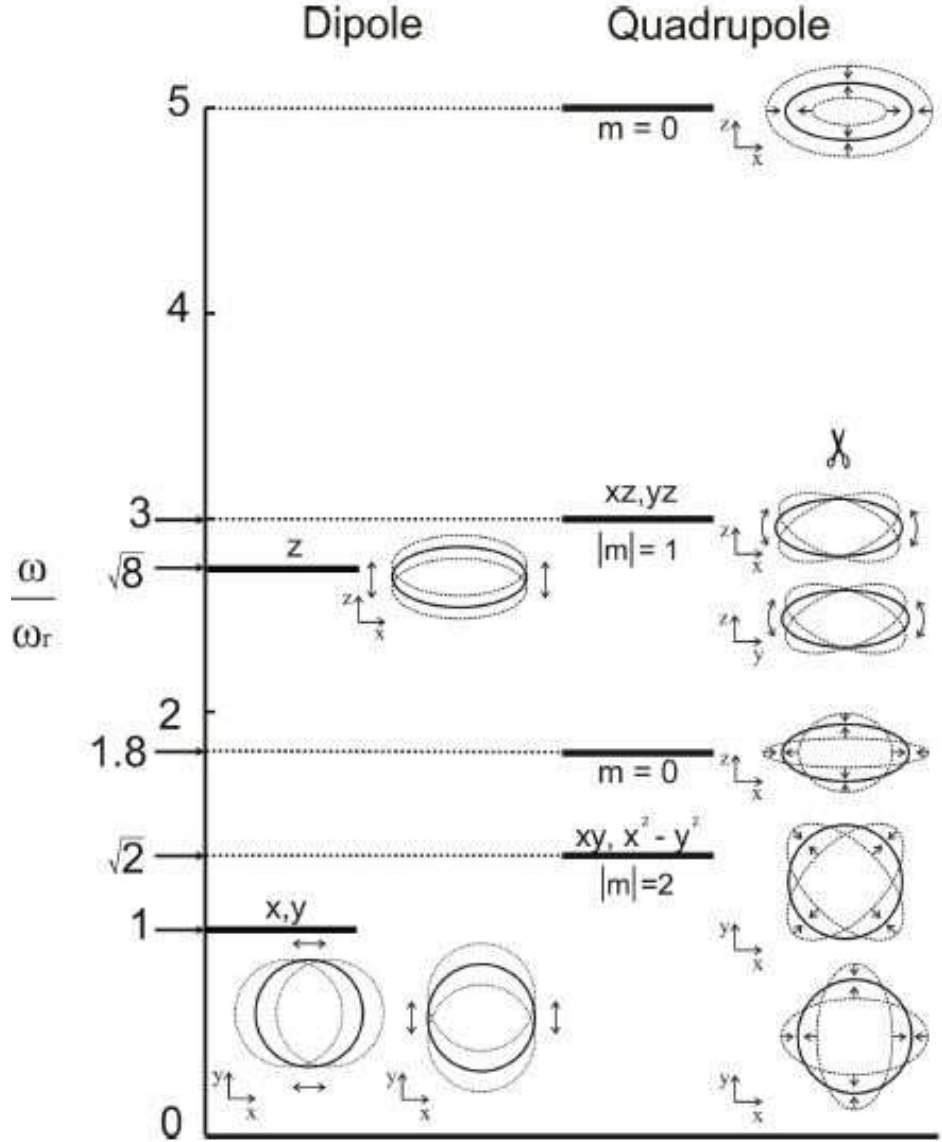}
\end{center}
\begin{flushleft}
\caption{\label{fig:CollectiveExcitationSpec} The low energy
collective oscillation mode spectrum of a trapped BEC in a TOP trap
where $\omega_{z}/\omega_{r} = \sqrt{8}$: the three dipole modes are
two degenerate modes at $\omega_{r}$ and one at $\omega_{z}$;  the
six quadrupole modes are the $m = 0$ low-lying and high-lying modes,
the doubly degenerate $\vert m\vert = 2 $ modes and the $\vert
m\vert = 1 $ or scissors modes. Figure drawn by E. Nugent, Oxford. }
\end{flushleft}
\end{figure}

Collective oscillation modes are classified according to the
symmetries of the trapping potential. In a spherical potential both
angular momentum and its axial component are conserved quantities
and solutions have the form $ n' = P_{\ell}^{(2
n_{r})}(r)r^{\ell}Y_{\ell m}(\theta, \phi)$ where $\ell$ is the
orbital angular momentum quantum number, $m$ is the magnetic quantum
number, $P_{\ell}^{(2 n_{r})}(r)$ are polynomials of degree $2n_r$
where $n_{r}$ is the number of nodes of the radial wave function,
and $Y_{\ell m}(\theta, \phi)$ are spherical harmonics. The
corresponding frequencies for a spherical harmonic potential are
$\omega(n_{r},\ell) = \omega_0(2 n_{r}^{2} + 2n_{r}\ell + 3 n_{r} +
\ell)^{1/2}$, where $\omega_0$ is the oscillation frequency of atoms
in the trap \cite{Pita03}. For $n_{r} = 0$, $\ell = 1$ the motion is
a simple dipole oscillation, or sloshing back and forth of the
cloud; in a harmonic trap this centre-of-mass motion
of the many-body system is decoupled from its internal degrees of freedom. 
Thus the dipole modes have frequencies exactly equal to those of the
oscillations of individual atoms in the trap which facilitates
calibration of experiments. The modes with orbital angular momentum
quantum number $\ell =2$ (and $n_{r} = 0$) are quadrupole modes and
in a spherical potential the oscillation frequency is
$\sqrt{2}\omega_{0}$ for all five values of $m$. The mode with
$n_{r} = 1$, $\ell = 0$ is known as the ``breathing mode" in which
there is expansion and contraction along the radial direction
changing the volume of the cloud but not its shape and this mode has
a frequency of $\sqrt{5}\omega_0$. Experiments are rarely performed
in a spherical trap but in many cases the potential does have a
cylindrical symmetry about the $z$-axis ($\omega_x =\omega_y
=\omega_\perp$) so that $m$ remains a good quantum number (but not
$\ell$), and two of the three dipole modes remain degenerate. In an
axisymmetric potential the $m=0$ modes with $\ell = 0$ and $\ell =
2$ are mixed to give modes for which the oscillations along the
axial and radial directions are in phase (higher frequency) and out
of phase (lower frequency), and the $\vert m \vert=2$ modes remain
degenerate (as do the $\vert m \vert =1$ modes). The frequency
spectrum of the modes and their shape are shown in Fig.\
\ref{fig:CollectiveExcitationSpec} for a time averaged orbiting
potential (TOP) trap with $\omega_{z}/\omega_{r} = \sqrt{8}$ (see Sec.~12.2 of~\cite{Pita03} for more general references).

Some of the first experiments on BEC measured the frequencies of the low-lying $m=0$ mode and an $m=2$ quadrupole mode and found good agreement with hydrodynamic theory when the number of atoms in the condensate is sufficiently large.
After this verification, considerable subsequent theoretical and experimental effort was concentrated on the so-called scissors modes with $\vert m \vert =1$ as described in Sec.~\ref{Sec_scissors}.
The scissors mode differs from the other collective oscillation modes
because it involves rotational motion about an axis (lying in the $xy$ plane)
as illustrated in Fig.\ \ref{fig:CollectiveExcitationSpec}. Thus it is sensitive to the irrotational flow
associated with superfluids.

Interactions give rise to a nonlinear term in the Schr{\"o}dinger
equation, Eq.\ (\ref{GPE}), and this nonlinearity leads to
a coupling between the collective excitations, e.g.\  
second harmonic generation was observed 
when the trapping potential was adjusted so that the frequency of a
high-lying mode was close to the second harmonic of the low-lying
$m=0$ mode \cite{Hechenblaikner2000}. When the BEC was driven at the
frequency of the lower mode there was a large transfer of energy to
the mode at twice the frequency.

The first experiments on collective oscillations were performed with
single component Bose gases in three-dimensional harmonic trapping
potentials, which is by far the simplest situation to analyze. More
recently there has been work on two-component gases, Fermi gases and
in complex and low dimensional trapping potentials. The collective
oscillation frequencies of a BEC do not depend on the strength of
the interactions (number of atoms) in a harmonic trap for reasons
explained above, but they do in a ring trap where the lowest-order
collective modes are quantized flows around the ring
\cite{Nugent2003}.  A very special theoretical case is a
two-dimensional interacting quantum gas confined in a harmonic
potential which has a breathing mode at the universal frequency of
precisely $2\omega_\perp$, twice the radial oscillation frequency,
because of underlying symmetry of quantum mechanical Hamiltonian
\cite{Pita97}. A wealth of new physics beyond mean field theory
occurs in these systems and collective oscillations are an important
way of uncovering new effects.


\section{Dipolar condensates}\label{dip}

It is helpful to start with a brief review of electric and magnetic dipoles and their interactions (Ref.~\cite{Laha09} provides a good general review of these intriguing systems).  In SI units, an elementary electric dipole moment $\bm p$ has dimensions of charge-length (namely C\,m) with the electrostatic potential
\begin{equation}\label{Phid}
\Phi(\bm r) =\frac{ \bm p\cdot \bm r}{4\pi \epsilon_0 r^3}.
\end{equation}
where $\epsilon_0$  is the permittivity of the vacuum. 
A combination of the dipole electrostatic field $\bm E(\bm r) = -\bm \nabla \Phi(\bm r)$ and the energy $U = -\bm p\cdot \bm E(\bm r)$ of an {\em external} dipole in this field yields the interaction energy $U_{dd}$ of two dipoles $\bm p_1$ and $\bm p_2$
\begin{equation}\label{Edd}
U_{\rm dd} = \frac{\bm p_1\cdot\bm p_2 - 3(\hat r_{12}\cdot \bm p_1)(\hat r_{12}\cdot \bm p_2)}{4\pi \epsilon_0 \,r_{12}^3},
\end{equation}
where $\bm r_{12} = \bm r_1-\bm r_2$ is the vector joining the two dipoles.  The most notable features are the rather complicated angular dependence and the inverse-cube dependence on the separation  (a long-range interaction, in contrast to the usual short-range interaction $g\delta(\bm r)$ discussed in Sec.~3).

To simplify this interaction energy, it is common to orient the dipoles with an extra  applied electric field along $\hat z$, in which case Eq.~(\ref{Edd}) has the simpler form
\begin{equation}\label{Eorient}
U_{\rm dd} = \frac{C_{\rm dd}}{4\pi } \frac{1-3\cos^2\theta}{r_{12}^3},
\end{equation}
   where $C_{\rm dd} = p^2/\epsilon_0$ (assuming identical atomic dipoles) and  $\theta$ is the angle between the vector $\bm r_{12}$ and the $\bm{\hat z}$ axis.       Note the strong angular dependence that is proportional to the (negative of the) Legendre polynomial $P_2(\cos\theta)= (3\cos^2\theta-1)/2$:  two parallel dipoles aligned along the $\bm{\hat z}$ axis (head-to-tail) have negative (attractive) energy, whereas two parallel dipoles side by side have a positive (repulsive) energy. It is easy to see that the dimensional coupling constant $C_{\rm dd}$ has dimensions of energy-volume, exactly the same as the contact interaction constant $g$ from Sec.~3.

A similar but considerably more intricate argument yields the  identical expression (\ref{Eorient}) for the energy of oriented magnetic dipoles $\bm m$, where $\bm m$ has dimensions of current-area (A\,m$^2$).  The only difference is the definition of the magnetic-dipole  dimensional coupling constant $C_{\rm dd} = \mu_0 m^2$, where $\mu_0$ is the permeability of the vacuum. 
Note that, by definition,  $\epsilon_0\mu_0 = c^{-2}$.
Compared to an elementary electric dipole moment $p$,  an elementary magnetic moment $m$  has the same dimension as $cp$.  Furthermore,  a typical valence electron in an atom has orbital speed  of order $\alpha c$, where $  \alpha = e^2/(4\pi \epsilon_0 \hbar c) \approx 1/137$ is the fine-structure constant.  The atomic magnetic moment arises from electronic motion with velocity $v$, leading to the typical
order-of-magnitude estimate $m\sim vp$, with $v/c \sim \alpha$.  Thus the interaction energy of magnetic dipoles is  smaller by a factor of $\alpha^2 \sim 10^{-4}$ compared to that for electric dipoles~\cite{Laha09}.

One approach to parametrize the dipole-dipole interaction  is to introduce a corresponding dipolar length $a_{\rm dd} = C_{\rm dd} M/(12\pi \hbar^2)$.  For many purposes, the $s$-wave scattering length $a$  provides the relevant comparison, with the dimensionless ratio $\epsilon_{\rm dd} \equiv a_{\rm dd}/a$.  For $^{87}$Rb,  this ratio $ a_{\rm dd}/a \approx 0.007$ is small so that dipolar effects are frequently negligible (but see Sec.~\ref{spinor} for an example where the dipolar energy is crucial).

Equivalently, this same dimensionless ratio becomes $\epsilon_{\rm dd} =C_{\rm dd}/3g$~\cite{Laha09}, where $g$  is   the contact coupling constant.  At present, all experimental studies of dipolar effects have relied on the particular atomic species $^{52}$Cr that has a large intrinsic magnetic dipole moment $m \approx 6 \mu_B$, where $\mu_B = e\hbar/2m_e$  is the Bohr magneton.  Even in this most favorable case of $^{52}$Cr, the dimensionless ratio is only $\epsilon_{\rm dd} \approx 0.16$.  The Stuttgart group has created a BEC of $^{52}$Cr~\cite{Grie05}, which has allowed many detailed experimental studies. More recently, Lev and colleagues have made a BEC of $^{164}$Dy, which has an even larger atomic magnetic moment $\mu \approx 10 \ mu_B$~\cite{Lu11}

Recently, powerful laser techniques have created bosonic heteronuclear polar  molecules such as $^{40}$K$^{87}$Rb~\cite{Ospe08,Ni08}  that have large electric dipole moments with $\epsilon_{\rm dd} \approx 20$~\cite{Laha09}. The phase-space density required for a quantum degenerate gas of polar molecules has not yet been achieved, but strenuous efforts are currently being made in several experiments, and this is an extremely promising area for the future work. Such systems will qualitatively alter the study of ultracold quantum matter and offer many new possibilities.  Reference~\cite{Carr09} provides a recent review of such cold dipolar gases.

In the context of the Gross-Pitaevskii description,  the addition of the long-range dipole-dipole interaction potential significantly affects the total interaction energy [the third term in Eq.~(\ref{EGP})].  Specifically, the two-particle potential now becomes $g\delta(\bm r-\bm r') + U_{\rm dd}(\bm r-\bm r')$, leading to the more complicated contribution to the GP energy functional
\begin{equation}\label{Eint}
E_{\rm int}[\Psi]  ={ \textstyle{ \frac{1}{2}}} \int dV\,  \Psi^*(\bm r)\Psi^*(\bm r') \left[ g\delta(\bm r-\bm r') + U_{\rm dd}(\bm r-\bm r')\right] \Psi(\bm r')\Psi(\bm r) .
\end{equation}
Here, the first term leads to the local Hartree potential $V_{H} = g|\Psi(\bm r)|^2$, but the second dipole term is intrinsically nonlocal and couples the total condensate densities at the two different positions.  This second term significantly complicates the GP equation, and it is convenient to introduce  an additional dipole-dipole interaction potential
\begin{equation}\label{Vdd}
V_{\rm dd}(\bm r,t) = \int dV'\,U_{\rm dd}(\bm r-\bm r')\, |\Psi(\bm r',t)|^2,
\end{equation}
that acts in addition to the Hartree potential $V_H = g|\Psi|^2$.
The resulting  time-dependent GP equation
\begin{equation}\label{TDGPdd}
i\hbar \frac{\partial \Psi}{\partial t} = \left(-\frac{\hbar^2 \nabla^2}{2M} + V_{\rm tr} + g|\Psi|^2 + V_{\rm dd}\right) \Psi.
\end{equation}
now contains the nonlocal integral contribution $V_{\rm dd}$.  For the stationary case, the left-hand side of (\ref{TDGPdd}) simply reduces to $\mu \Psi$.

\begin{figure}[h]
\includegraphics[width=5in]{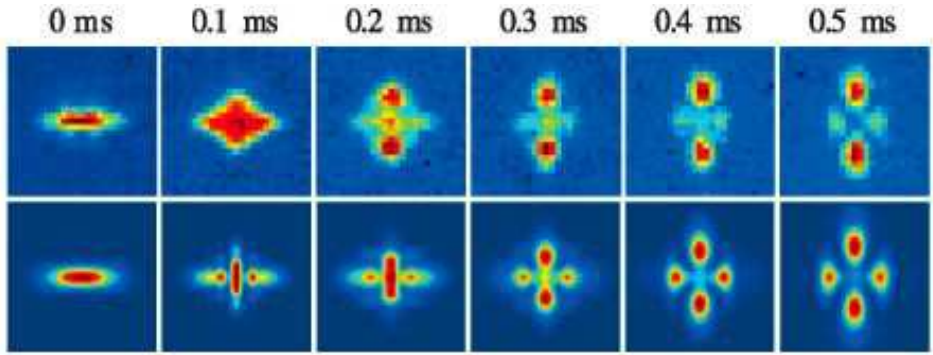}
  \caption{\label{dipole} Top row:  Experimental images of a dipolar condensate after collapse and explosion following a sequence of holding times for release from the trap (the system is axisymmetric around the horizontal axis). Bottom row:  Results of numerical simulation of the collapse dynamics with no adjustable parameters.  The field of view is  130 $\mu$m $\times\  130\ \mu$m. Reprinted with permission of the authors~\cite{Laha08} and the American Physical Society.}
  \end{figure}

Based on the form of Eq.~(\ref{Eorient}), a cigar-shaped pure dipolar condensate should collapse, whereas a disk-shaped pure dipolar condensate should be stable.  The repulsive contact interaction acts to counteract the collapse, and variational studies have confirmed these qualitative ideas, but the details become rather complicated~\cite{Laha09}.  Assume an axisymmetric trap with a particular aspect ratio (this depends on the trap frequencies $\omega_z$ and $\omega_\perp$).  Reference~\cite{Koch08} used the BEC of $^{52}$Cr to study the behavior of the condensate for various aspect ratios of the external trap.  Most importantly, they also used a Feshbach resonance to  control the $s$-wave scattering length and confirmed the theoretical picture.  For example, using a Feshbach resonance to tune the scattering length to zero, they demonstrated the stability of a pancake-shaped pure dipolar gas.  Subsequently, Ref.~\cite{Laha08} investigated the collapse dynamics of the dipolar condensate when the $s$-wave scattering length is too small to support the system.       As might be expected from the $P_2(\cos\theta)$ form of the oriented dipole-dipole interaction Eq.\ (\ref{Eorient}), Fig.~\ref{dipole} shows that the collapse dynamics leads to a $d$-wave symmetric implosion (the system is axisymmetric around the horizontal axis).

\section{Mixtures}\label{Sec_mixtures}

This section and the next consider multicomponent BECs, but there are two very different types.  The first is a mixture of different hyperfine states that are confined in a magnetic trap.  The second is a ``spinor condensate'' that involves all the magnetic sublevels of a single hyperfine state.  Since many of the sublevels cannot be magnetically trapped, this system requires an optical trap that can confine all the magnetic sublevels.

For definiteness, we consider the particular species $^{87}$Rb that has been used in many of the experiments with mixtures.  The nucleus has spin $3/2$ and the single $s$-state valence electron with spin $1/2$ leads to two hyperfine states.   The lower one has  $F =1$ and the higher one has $F=2$, separated by an energy with an equivalent frequency of  $6.8$ GHz.

In a magnetic trap,  among the $F=1$ manifold, only the state $|1,-1\rangle$ can be trapped (using the standard notation $|F,m_F\rangle$)   whereas the $F=2$  manifold  has two  states $|2,1\rangle$ and $|2,2\rangle$ that can be trapped.  Usually experiments focus on the pair $|1\rangle \equiv |1,-1\rangle$ and
 $|2\rangle \equiv |2,1\rangle$ since they have essentially the same magnetic moments and thus behave similarly in a magnetic field (which means that they experience similar trap potentials).

\subsection{interacting two-component mixtures}
The simplest case of a mixture is  two dilute Bose gases with order parameters $\Psi_1$ and $\Psi_2$ and coupling constants $g_{11}$ and $g_{22}$ for self-interaction and $g_{12}= g_{21}$ for mutual interaction.  The principal new feature is that the GP equation for (say) $\Psi_1$ now includes an additional Hartree term $V_{H12}(\bm r)=  g_{12} |\Psi_2(\bm r)|^2$ for the mutual interaction with species 2
\begin{equation}\label{GP1}
i\hbar\frac{\partial \Psi_1(\bm r,t)}{\partial t} = \left[-\frac{\hbar^2\nabla^2}{2M} + V_{\rm tr1}(\bm r) + g_{11}|\Psi_1(\bm r,t)|^2+ g_{12} |\Psi_2(\bm r,t)|^2 \right ]\Psi_1(\bm r,t),
\end{equation}
and similarly for the GP equation for $\Psi_2$.

Suppose for simplicity that the two condensates are uniform.  It is not difficult to see that the conditions for miscibility are~\cite[Sec.~12.1]{Peth08}
\begin{equation}\label{stab}
g_{11} >0,\quad g_{22} > 0,\quad\hbox{and}\quad g_{11}g_{22} > g_{12}^2.
\end{equation}
The first and second conditions ensure that each pure condensate is stable against collapse (as is familiar for a single-component condensate).  In contrast, the third condition ensures that the overlapping mixture is stable against phase separation.  If this latter condition is violated, then the uniform two-component system will phase separate, and the long-wavelength collective modes of the coupled uniform system will have imaginary frequencies.

The situation in a trap is somewhat different, since the harmonic trap acts to confine the two components.  Unless the self-interaction constants are essentially equal, one component  will  form a denser core, surrounded by (and partially overlapping with) the other less-dense component   that  experiences the inward pressure of the trap and  outward pressure of the inner core (analogous to density stratification in an inhomogeneous  self-gravitating body).   

As explained below, external electromagnetic fields allow a convenient transformation between these hyperfine species, for example a sudden transition  from $|1\rangle$ to $|2\rangle$.  Reference~\cite{Matt98} modeled this behavior with a single-component time-dependent GP equation with a time-dependent scattering length $a(t)$ that changed from $a_{11}$ to $a_{22}$ discontinuously at $t = 0$.   After a variable time-delay, the trap was turned off,  allowing the condensate to expand.  Initially, the condensate shrank, followed by compressional oscillations.  This behavior is understandable because $a_{11} >a_{22}$ for these particular states.  Thus the initial density profile for $|1\rangle$ could no longer sustain the equilibrium radius once the species transformed to $|2\rangle$, experiencing radial shrinkage and radial oscillations.  A fit to the observations indicated that $a_{11}/a_{22} \approx 1.06$.

\subsection{electromagnetic coupling between two hyperfine states}

The previous model in (\ref{GP1}) assumes  that the two components interact only through the density (the mean-field Hartree terms) that contain no phase information.  Specifically, each condensate has the representation $\Psi_j(\bm r) =|\Psi_j(\bm r)| \exp[iS_j(\bm r)]$.  Each order parameter is single-valued when $\bm r$ executes a closed path at any given instant, so that the phase is $2\pi$-periodic for any such closed path.  As noted in connection with superfluid $^4$He, the circulation for each component is therefore  quantized in units of  $h/M$. Technically, each complex one-component order parameter has $U(1)$ symmetry. Far below the superfluid transition temperature, the quantized circulation ``explains'' the existence of persistent currents  because any fluctuation that can change the circulation is very improbable, reflecting the topological charge associated with the $2\pi$ periodicity.

The situation is very different when external electromagnetic fields couple the two components, leading to off-diagonal terms in the combined GP equations.  The resulting  total order parameter no longer has two separate complex scalar functions [each with $U(1)$ symmetry];  instead  it  now has two components with intrinsic coupling,  qualitatively changing its character into a single $SU(2)$ structure.  For elementary discussions of this essential  model system, see, for example,~\cite[Chaps.\ 7-11]{Feyn65} and~\cite[Sec.~4.4]{Grif05}. The dynamics of this spin-$\frac{1}{2}$ becomes similar to that familiar from nuclear magnetic resonance, obeying what are here called the ``optical Bloch equations.''  In the present case, the transformation from $|1\rangle$ to $|2\rangle$ is analogous to a $\pi$ pulse  that rotates a spin up into a spin down.  If the pulse is twice as long, the resulting $2\pi$ rotation reproduces the initial $|1\rangle$ state apart from an overall phase.    Experiments~\cite{Matt98} verify the periodic transfer between the two states as the length of the applied pulse varies (each hyperfine state has a slightly distinct resonant frequency, so that each can be imaged separately).

This electromagnetic coupling has the remarkable feature of allowing a direct measurement of the relative phase between the two components, as becomes clear from the dynamical equations of the coupled system~\cite{Hall98}
\begin{equation}\label{coupled}
i\hbar\frac{\partial}{\partial t}
\left(
\begin{matrix}\Psi_1\\ \Psi_2
\end{matrix}
\right)
 \approx
 \left(
 \begin{matrix} {\cal T} + V_{\rm tr1} + V_{H1} + V_{H12} & \frac{1}{2}\,\hbar\Omega(t)\exp(i\omega_{\rm rf}t)\\
\frac{1}{2}\,  \hbar\Omega(t)\exp(-i\omega_{\rm rf}t) & {\cal T} + V_{\rm tr2} + V_{H2} + V_{H12}
  \end{matrix}\right)
  \left(\begin{matrix} \Psi_1 \\\Psi_2
  \end{matrix} \right),
\end{equation}
where $\cal T$ denotes the kinetic energy, $\omega_{\rm rf}$ is the frequency separating the two hyperfine states, and $\Omega(t)$ reflects the two-photon electromagnetic coupling between the two hyperfine states.  It is known as the ``Rabi'' frequency and depends on the strength of the electromagnetic coupling field (the time dependence here arises because it can be turned on and off, but more general situations can and do arise).  Transforming to the rotating frame eliminates the explicit  rf time dependence.

The experiment starts with population in $|1\rangle$ and applies a $\pi/2$ pulse that creates a linear superposition of the two states, then waits a variable time and applies a second $\pi/2$ pulse.  The trap is  turned off, expanding the condensate.   Resulting images of either component show oscillations depending on the length of the variable delay, which provide a measure of the relative phase.

This and other experiments indicate that the electromagnetic coupling has the remarkable property of changing the topology of this two-component system.  When the coupling is off, they are simply separate $U(1)$ complex scalar order parameters, each with its own phase angle and corresponding  quantized circulation (each has the topology of a cylinder).  When the coupling is on, they instead form a single coupled $SU(2)$ system with two angles on the polar sphere and  no topological quantization (the topology is now that of a sphere).   The JILA group studied this system in considerable detail, and its topology proved essential in the first experimental creation of a  quantized vortex in a dilute BEC~\cite{Matt99}, as discussed in Sec.~\ref{rot}.

\section{Spinor condensates}\label{spinor}

The previous section focused on mixtures that usually involve two  different $F$  hyperfine manifolds in a magnetic trap.  In this situation, only the weak-field seeking states are relevant, since the other states no longer remain confined.  In contrast, a spinor condensate involves a single hyperfine manifold, but it uses an optical trap that retains all the $m_F$ magnetic states (the simplest and  usual case is $F=1$ so that the $m_F$ levels are $+1$, 0, and $-1$).  See Chap.~1 for an introductory discussion of these traps.

\subsection{spinor condensates:  special case of $F=1$}

The possibility of trapping all three magnetic sublevels in an optical trap rapidly led to the study of spin-1 Bose-Einstein condensates (as well as more general cases)~\cite{Ho98,Ohmi98}.  In contrast to the previous mixture of distinct species, the rotational invariance of the interactions between two spin-1 atoms now  leads to special restrictions.  Here, the macroscopic order parameter is a three-component vector  (written as a transpose for convenience)
\begin{equation}\label{vector}
\bm \Psi^T = \left(\begin{matrix}
\Psi_1&
\Psi_0&
\Psi_{-1}
\end{matrix}\right)
\end{equation}
involving the three $m_F$ states.  Only $s$-wave scattering is relevant in the present low-energy limit, and the interaction between the atoms has the familiar contact form
\begin{equation}\label{contact}
V_{\rm int} (\bm r_1-\bm r_2) = \delta(\bm r_1-\bm r_2) \sum_F\frac{4\pi \hbar^2 a_F}{M}\, {\cal P}_F,
\end{equation}
where $F$ is the magnitude of the total hyperfine spin of the two atoms $\bm F = \bm F_1+\bm F_2$ and ${\cal P}_F$ is the projection operator onto the appropriate value of the total $F$.

For bosons each with hyperfine spin $F_j=1$, the only allowed total values are $F=0$ and $F=2$, with two scattering lengths $a_0$ and $a_2$.  The interaction potential can then be rewritten in the equivalent form
\begin{equation}\label{contact1}
V_{\rm int} (\bm r_1-\bm r_2) = \delta(\bm r_1- \bm r_2)  \left(g_0 + g_2\,\bm F_1\cdot\bm F_2\right),
\end{equation}
where the effective interaction constants are given by
\begin{equation}\label{effint}
g_0 = \frac{4\pi \hbar^2}{M}\,\frac{2a_2+ a_0}{3},\quad g_2 = \frac{4\pi \hbar^2}{M}\,\frac{a_2- a_0}{3}.
\end{equation}
In practice, the two scattering lengths are roughly comparable, so that $|g_2|\ll g_0$.  Specifically, $g_2$ is small and positive for $^{23}$Na, but it is small and negative for $^{87}$Rb, which has crucial practical consequences for the distinct behavior of these two atomic species.

Ho~\cite{Ho98} introduced an effective energy functional, writing the spinor order parameter in the form   $\Psi_\alpha(\bm r) = \sqrt{n(\bm r)}\,\zeta_\alpha(\bm r)$, where the spinor index runs over the values $1, 0, -1$, $n(\bm r)$ is the common density for all three components, and $\zeta_\alpha(\bm r)$ is a normalized spinor with $\bm \zeta^\dagger\cdot \bm \zeta = 1$.  The ground state follows by minimizing the energy with fixed total particle number (enforced with a chemical potential), leading to an effective energy functional $K = E -\mu N$
\begin{equation}\label{Espinor}
K = \int dV \left[ \frac{\hbar^2 }{2M} \left|\bm \nabla \sqrt n\right|^2 +\frac{\hbar^2}{2M} \left|\bm \nabla \bm \zeta\right| ^2 -n[\mu - V_{\rm tr}(\bm r)] + \frac{n^2}{2}\left(g_0 + g_2\langle\bm F\rangle^2\right)  \right].
\end{equation}
Here the trap is taken as independent of the hyperfine state, and $\mu$ determines the total number of atoms.

Apart from the gradient terms, the ground-state spinor $\zeta_\alpha$ follows by minimizing the spin-dependent part of the energy $\frac{1}{2} n^2 g_2\langle \bm F\rangle ^2$, where $\langle \bm F\rangle  = \sum_{\alpha\beta} \zeta_\alpha^\dagger \bm F_{\alpha \beta}\zeta_\beta$ is the expectation value of the appropriate spin-1 matrices~\cite[Sec.~12.2.1]{Peth08}.  If $g_2$ is positive (as for $^{23}$Na), then the minimum occurs for $|\langle \bm F\rangle | = 0$.  Such states are called ``polar'' and they are those obtained by spatial rotations of the hyperfine state $|m_F = 0\rangle$.  If $g_2$ is negative (as for $^{87}$Rb), then the minimum occurs by maximizing the average spin with $|\langle \bm F\rangle |~=~1$. These states are called ``ferromagnetic'' and they are those obtained by spatial rotation of the hyperfine state $|m_F = 1\rangle$.

\subsection{experimental studies of spinor condensates}

Soon after the optical trapping of a BEC of $^{23}$Na atoms, the MIT group made a detailed experimental study of these fascinating  spin-1 spinor systems, where the  spin-dependent  interaction constant $g_2$ in (\ref{effint})  is small and positive,  favoring polar configurations with $\langle \bm F\rangle = 0$.  For definiteness, we  focus on the formation of ground-state spin domains in an external magnetic field~\cite{Sten98}.  The resulting spin structures can be either miscible or immiscible depending on the applied magnetic field and on which of the $m_F$ components are occupied.  They imaged the  spin domains  by time-of-flight expansion followed by a Stern-Gerlach separation of the various spin components in an inhomogeneous magnetic field. Reference~\cite{Stam00} comprehensively reviews this and related experiments on the polar spinor condensate $^{23}$Na.

Recently, Stamper-Kurn's group at Berkeley has  studied the very different situation in $^{87}$Rb, where the relevant spin-dependent  coupling constant $g_2$ in (\ref{effint}) is small and negative, favoring ferromagnetic spinor structures with $|\langle \bm F\rangle| = 1$~\cite{Veng08}.   Here, the relevant scattering length is not   the usual $s$-wave $a$, but the much smaller spin-dependent quantity $\Delta a = (a_2-a_0)/3 $ that follows from Eq.~(\ref{effint}).  Consequently,  the usually small dipolar length $a_{\rm dd}$ now becomes crucial.  Specifically, the dimensionless ratio $|a_{\rm dd}/\Delta a|\approx 0.4$ is no longer small, which leads to many intriguing new phenomena involving the dipolar energy (effectively a dipolar quantum fluid~\cite{Veng08}).

\begin{figure}[ht]
\includegraphics[width=5in]{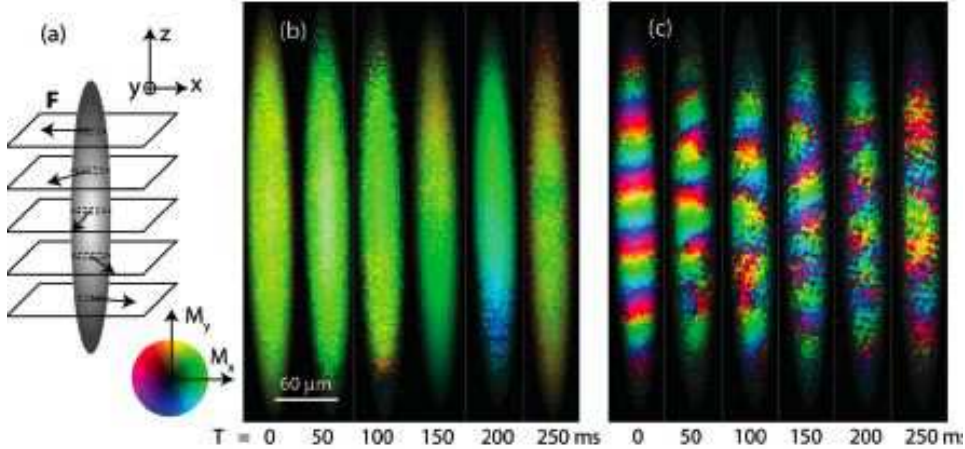}
  \caption{\label{helix}  Spontaneous dissolution of helical textures in a quantum degenerate $^{87}$Rb spinor Bose gas.  A magnetic field gradient   prepares transversely magnetized  (b) uniform or (a), (c) helical magnetic textures.  The transverse magnetization after a variable time  of free evolution is imaged in the $xz$ plane, with orientation indicated by the hue and amplitude brightness.  (b) Uniform texture remains homogeneous for long evolution times, whereas (c) with initial  pitch $\lambda = 60 \ \mu$m dissolves over $\sim $200 ms, leading to a sharply modulated texture. Reprinted with permission of the authors~\cite{Veng08} and the American Physical Society.}
  \end{figure}
The experiment involved an asymmetric triaxial trap with a thin elongated elliptical two-dimensional TF condensate that initially had a helical spin texture with a variable pitch (periodicity length $\lambda$ between 50 and 150 $\mu$m).  This helical spin texture evolved  for a variable time, and the vector magnetization was then measured nondestructively.  For comparison, they also prepared similar uniform samples with much larger pitch ($\lambda \gg R_z$).  Figure~\ref{helix} shows the evolution of these  initial  textures (a), including both (b) uniform  and (c) helical  structures.  For (c), note the small-scale structure of the magnetization of order $10 \ \mu$m, much smaller than the initial helical pitch. Several experimental  tests associate this short-range modulation  with the dipolar interaction.

\section{Rotating Bose gases and quantized vortices}\label{rot}

The superfluid properties of liquid helium at low temperatures
(below the lambda-point) were apparent soon after liquefaction was
achieved at the beginning of the 20th Century, in particular the
absence of viscosity that allows the liquid to leak through small
channels. The properties of superfluid helium were explained in an
intuitive phenomenological way by the two-fluid model (of Landau and
others) outlined in Sec.~\ref{2fl}; although there is a connection
with Bose-Einstein condensation, it has not been central to
understanding the superfluid helium as can be seen by comparing the
physics discussed in Sec.~\ref{quasi} with that in
\ref{hydroequations}.

 The study of
BEC in dilute atomic gases developed in a very different way. The
first experimental breakthroughs confirmed the prediction of Bose
and Einstein and found that the critical temperature was that predicted by statistical
mechanics (within a few percent). 
About four years after the first BEC was created, experiments
started to demonstrate  the striking superfluid properties that
distinguish it from a classical fluid. This section mainly concerns
those methods that use rotation to probe the quantum properties. For
completeness, however, we  mention the work on the flow of the
quantum fluid around a strongly repulsive potential created by a
focused blue-detuned laser beam. Displacing this laser beam back and
forth acts like a `macroscopic object' moving inside the cloud of
atoms;  it was found that there is a critical velocity [defined in
Eq.~(\ref{vcBog})] below which there is no observable dissipation of
energy.

\subsection{the scissors mode}\label{Sec_scissors}

Some of the first evidence for the superfluidity of dilute atomic
gases in the BEC regime came from the quantitative measurements of
the so-called scissors mode which derives its name from nuclear
physics where there is an excitation of the protons and neutrons in
deformed nuclei (as explained below) \cite{Guery-Odelin1999}.  In the case of BEC the cloud
of ultracold atoms is non-spherical because of asymmetry in the
trapping potential, i.e.\ the three frequencies in Eq.~(\ref{Vharm}) are not
all the same (which is almost always true in practice), so that the
density contours are deformed into ellipses. When the potential is
suddenly rotated through a small angle about a symmetry axis, the
deformed density distribution finds itself offset from its
equilibrium position~\cite{Marago2000} and therefore starts rotating to and fro
through a small angle without changing shape corresponding to the modes
labelled $|m=1|$ in Fig.\ \ref{fig:CollectiveExcitationSpec}. Superfluids behave in a distinctive way
in rotating systems because of their irrotation flow ($\bm\nabla \bm \times \bm v_s = 0$ as discussed in Sec.~\ref{2fl}); this leads to a
reduced moment of inertia as compared to the classical value for a
`solid-body' with the same density distribution.

In trapped three-dimensional ultracold gases the superfluid and BEC
fraction are considered to be the same (the condensate fraction is small in
helium, as discussed in Sec.~\ref{Sec_condensate}). Two-dimensional systems, however, can be completely
superfluid even when there is no BEC (see Sec.~\ref{Sec_nucleation}), and there is great interest in investigating the relationship between superfluidity and BEC.

As mentioned above the scissors mode experiments with ultracold
gases were suggested \cite{Guery-Odelin1999} based on an analogy with
groundbreaking work on superfluidity of nuclear matter. The scissors
mode in strongly deformed nuclei was first suggested in a model
where the protons and neutrons are assumed to be two distinct
deformed quantum fluids, e.g.\ two ellipsoidal distributions with a
common center. These deformed distributions rotate relative to each other around a common axis
(two-rotor model) like the blades of a pair of scissors about the pivot.

\subsection{the nucleation of vortices}\label{Sec_nucleation}

Perhaps the most obvious way to create a vortex in a fluid is rotating the container as in experiments with superfluid helium, or
using laser beams moving with a circular motion to stir the system like a spoon in a cup of tea. However the first vortex created in a BEC of
ultracold Rb atoms was achieved at JILA by an ingenious scheme
which directly controlled the phase of wave function, in a manner
far removed from any previous work (on helium) \cite{Anderson2000}. That experiment used
a mixture of ultracold atoms in two different magnetic states as described in Sec.~\ref{Sec_mixtures}. By driving a two-photon transition between the states (with a combination of microwave and RF radiation), a
ring of atoms was created which had a phase winding of $2\pi$
corresponding to a vortex state, like a persistent flow
with one unit of angular momentum (and further work on persistent
flow is discussed below). In this scheme the center of the ring was
filled with atoms in a different Zeeman sub-state, but these atoms
were selectively removed to leave a cloud of atoms containing a more
conventional vortex (and much of the rich physics of vortices and other topological structures in multi-component systems remains to be explored).

In principle this scheme is very flexible, but much of the subsequent work has
 used experimental methods in which mechanical rotation
or stirring imparts angular moment to the condensate. Analogous to the ``rotating bucket'' experiment in superfluid $^4$He, ultracold atoms have been
trapped in a rotating potential whose contours of constant energy are elliptical (at a given instant of time).  However once a BEC has formed and established
long-range phase coherence there is significant hysteresis: the
condensate can remain in a (metastable) state without vortices
even when the confining potential rotates sufficiently fast that vortex states
have lower energy. The nucleation of vortices requires a way to
overcome the energy barrier. This has been achieved by driving the
system at a frequency that resonantly excites a quadrupole mode so
that deformation builds up until the cloud of atoms is sufficiently
perturbed that vortices can enter; vortices form in the
low-density outer reaches of the cloud and then move towards the
center.

 The characteristic core
size in the healing length $\xi$ given in Eq.~(\ref{xi}). The wave
function for a vortex has the form $\Psi(r,\phi) = f(r/\xi)
e^{i\phi}$, and a variational solution of the GP equation gives the
approximate amplitude as $f(x) = x/\sqrt{x^2 +2}$ which provides a
good fit at both large and small $x = r/\xi$ (when compared to more
accurate numerical solutions).

\begin{figure}[ht] 

\begin{center}
\includegraphics[scale=0.65]{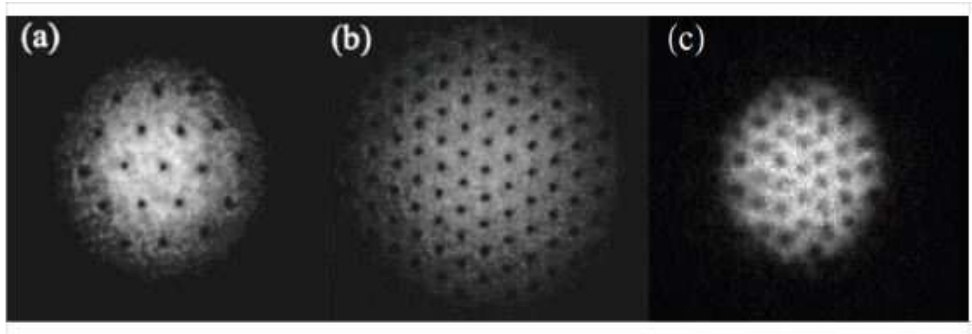}

\end{center}
%
 \caption{\label{fig:Codd} Images of expanded Bose-Einstein condensates of $^{87}$Rb atoms showing (a) small vortex array for slow rotation, and (b) large vortex array for rapid rotation.  Note the highly regular triangular form. (c) An array of vortices that has not settled down to a regular lattice as in the previous images; stirring or shaking such a system creates a tangle of vortex lines corresponding to a turbulent quantum fluid. Images (a) and (b) reprinted with permission of the authors~\cite{Codd04} and the American Physical Society.  Image (c) provided by R.\ Williams and S.\ Al-Assam \cite{williams2010}.}  
\end{figure}

Rather than starting with a static BEC and
trying to break up the established phase, an alternative strategy
for nucleating vortices is to spin up a thermal cloud of trapped atoms and
then cool them below the critical temperature in the rotating frame.
Arranging preferentially to  remove atoms with less than average
angular momentum during evaporation increases the rotation rate  as the
temperature decreases. This puts the system directly into the
equilibrium state for the given rotation rate, i.e.\ the quantum
fluid has the number of vortices corresponding to the rotation rate
of an equivalent classical fluid, see Fig.~\ref{fig:Codd} (as outlined in Sec.~\ref{circ}).

Generally the vortices are too small to be observed in a trapped BEC
and images such as Fig.~\ref{fig:Codd} are taken after a
time-of-flight expansion. Indeed for pancake-shaped condensates the vortices
expand proportionally more than the overall size of the cloud to
give extremely high contrast images. By observing along the axis of
rotation the cores of the vortices appear as dips in the density of
the cloud of atoms (in images recorded by absorption of resonant
laser light).

Recently another method has been developed that is closely related
to the original phase imprinting scheme \cite{williams2010}. A
Bose-Einstein condensate is loaded into a deep 2-D optical lattice
to give an array small condensates (containing hundreds of atoms at
each site); the phases of these separated groups of atoms can evolve
independently so that when the optical lattice potential is rotated,
about an axis perpendicular to the plane of the lattice, phase
differences build up corresponding to the velocity field in a
rotating system [see Eq.~(\ref{vs})]. Thus when the lattice depth is
reduced adiabatically to zero the small clouds join together with
relative phases that give vortices throughout the system (with no
hysteresis). 
Vortices rotating in the same direction repel
each other and they arrange themselves into a lowest energy
configuration that is a regular triangular array (or Abrikosov
lattice similar to that of vortices in superconductors), assuming
that there is some damping in the system.

In some situations, the regular arrangement of vortices aligned along the
rotation axis as in Figs.~\ref{fig:Codd} can be disrupted to
create a turbulent system. In superfluid $^4$He, the much smaller
healing length and larger sample size allows for many
vortices and quantum turbulence in the evolution of a tangle of
vortices has been studied. Another fascinating type of system is
ring vortices that resemble smoke rings in the condensate; these
have been created in experiments with ultracold atoms by producing a
sharp phase discontinuity in the BEC that evolves into a ring vortex
(albeit with less contrast than for vortices whose cores go right
though the cloud). In addition to studying the properties of
individual vortices, these systems allow the study of the collective
properties of ``vortex matter''. Interesting effects are predicted to
occur when the number of vortices becomes very large, corresponding
to very fast rotation, as described in the following section.

In addition to the work on BEC, vortices have been used to investigate superfluidity in quantum degenerate Fermi gases \cite{Zwierlein05}, which has connections with other systems such as superconductors, neutron stars etc. Cold atomic vapors have the important characteristic  that the strength of interactions between atoms can be precisely controlled over an enormous range (for both fermions, bosons and mixtures) thus making these systems a fantastic test bed for theoretical ideas.  The experiment of Zwierlein {\it et al.}\ at MIT \cite{Zwierlein05} used the fermionic isotope of lithium ($^6$Li). Adjusting the magnetic field experienced by the atoms to increase the strength of interactions (near a Fano-Feshbach resonance where the scattering length is resonantly enhanced) causes pairs of atoms to bind together to form long-range molecules, i.e.\ molecules in a very high-lying vibrational level of the molecular potential where the internuclear spacing is orders of magnitude greater than that of a Li$_2$ molecule in its ground vibrational level. These molecules are composite bosons and hence at low temperature they are in a BEC. A vortex lattice was created in this BEC by imparting angular momentum to the system by rotation, as in experiments with bosonic atoms. The binding energy was decreased (by changing the magnetic field) so that the ultracold molecules became weakly bound pairs of fermions closely analogous to the Cooper pairs (of electrons) in superconductors. This allowed a detailed study of the crossover from a BEC of molecules to a Bardeen-Cooper-Schrieffer superfluid of loosely bound pairs, and this particular experiment showed that superfluidity as evidenced by the existence of vortices persisted in this BEC-BCS cross-over.  The ability to scan the system from one regime to another is a truly remarkable feature of ultracold vapors.

The first observation of a vortex in an ultracold gas was a challenging experiment but nowadays vortices have been observed in a
great variety of ways; even in non-rotating systems where vortices can (sometimes) arise spontaneously after a
rapid quench, i.e.\  sudden cooling through the phase transition, or by merging Bose-Einstein condensed clouds that have different (random) phases.  Another method that has be successfully used to nucleate vortices spontaneously, without rotation or stirring, is by merging of multiple trapped BECs \cite{Scherer07}. In this experiment three BECs were formed that were separated from one another and then the potential barriers between them  were lowered so that they merged together. The independent, uncorrelated BECs had different (random) phases and sometimes there was sufficient phase winding for vortices to form.

The physics of the persistent flow of a superfluid
around a ring and a quantum vortex are closely linked. For a
ring-shaped cloud of BEC the atoms are at a fixed radius and therefore the
condition on circulation in Eq.~(\ref{circ1}) corresponds exactly to quantisation of
angular moment in units of $\hbar$. Note, however, that vortices with more
than one unit of circulation are energetically unstable, e.g.\ a vortex with two units rapidly breaks up into two vortices.
This has been used to measure angular momentum of flow around a ring
formed by the combination of a repulsive barrier (from a laser beam
with blue frequency detuning) passing through the center of a
trapping potential; slowly reducing the height of the central
barrier to zero puts the atoms into a harmonic trap where the ring-shaped cloud
rapidly separates in many vortices of unit circulation which can be
counted. Vortices with circulation greater than unity can be
energetically stable in anharmonic potentials, e.g.\ when there is a
positive quartic ($r^4$) term in addition to harmonic confinement
($r^2$). Such traps are of interest in the pursuit
of the quantum states that arise at very fast rotation rates as described below.

\


\subsection{the use of rotation for direct quantum simulation}

The rotation of neutral atoms can be considered as a way of
simulating the effect of a magnetic field $\bm{B}$ on charged
particles.
A simple illustration of this analogy can be seen by comparing the form of the Lorentz force on a charged particle in a magnetic field,
$q(\bm{v}\times\bm{B})$, and the Coriolis force on a neutral particle of mass $M$ in the rotating frame,
$2M(\bm{v}\times\mathbf{\Omega})$, suggesting a mapping
\begin{equation}
q\bm{B} = 2M\bf{\Omega}. \label{eqn:B_Omega}
\end{equation}
where we generally take  $\bm \Omega = \Omega\hat{\bm z} $. More rigorously one can consider the form of the Hamiltonian of a
single neutral particle in a 3D harmonic trapping potential in the
rotating frame,
\begin{eqnarray}
H_\Omega &=& H_0 - \bm{\Omega\cdot L}\\
&=& {\frac{1}{2M}}\left[{\bm{p}} - M({\bm{\Omega\times r}})\right]^2 +  \textstyle{\frac{1}{2}}  M(\omega_\perp^2-\Omega^2)(x^2+y^2) +  \textstyle{\frac{1}{2}} M\omega_z^2z^2.
\label{eqn:rot_hamilt}
\end{eqnarray}
It is apparent that the kinetic energy term is formally equivalent
to the usual gauge-invariant term $(\bm{p}-\bm{A})^2/2M$, with
$\bm{B}=\nabla\times\bm{A}$, that arises for a charged particle in a
magnetic field. (We take the effective charge as~$1$ so $\bm A$ has
the dimension of momentum~\cite{Dali10}.)  Note that the centripetal
acceleration which arises from rotation reduces the radial harmonic
trapping frequency, $\omega_\perp \rightarrow
(\omega_\perp^2-\Omega^2)^{1/2}$.  Thus in experimental conditions
where the rotation frequency $\Omega$ tends to the radial trapping
frequency $\omega_\perp$ radial trapping becomes very weak and the
atomic cloud expands to be a very thin pancake-shape. Under these
conditions the temperature of the quantum degenerate gas is so low
that $k_B T \ll \hbar \omega_z$ and the motion in this direction is
frozen out. This makes the Hamiltonian of Eq.\
(\ref{eqn:rot_hamilt}) almost equivalent to that of a
two-dimensional electron gas under a magnetic field (but with some
residual harmonic potential in the radial direction).  Interactions
have not been included in this description but typically the
chemical potential $\mu = gn(0)$ is small in this regime because of
the cloud has low density, $n(0)$, when it has spread out radially.
The close analogy between the physics of rotating neutral atoms and
electrons under a magnetic field has led to considerable interest in
the possibility of achieving strongly-correlated quantum Hall states
in a rapidly rotating atomic gas (for a recent review see
\cite{cooper2008}).

The use of rotation for simulating a gauge field with ultracold
atoms has a significant drawback, namely exotic strongly-correlated
states are predicted to occur in a Bose gas rotating very close to
the centrifugal limit, $\Omega\simeq\omega_\perp$, at which rotating
atoms are expelled from the trap.  There are considerable technical
challenges associated with rotating close to the centrifugal limit,
e.g.\  the harmonic trap must be very axisymmetric ($\omega_x$ very
close to $\omega_y$), and any static perturbations in the lab frame
lead to substantial heating. These issues have prevented fractional
quantum Hall effect physics from being achieved in a rotating Bose
gas, though impressive attempts have been made
\cite{schweikhard2004}. In order to circumvent the pragmatic
problems associated with rotation there has been a development of
alternative methods for creating synthetic gauge fields for
ultracold atoms~\cite{Dali10}, as described in the following
section.



\section{Synthetic (artificial) gauge fields---vortices without rotation}\label{syn}


This particular  vector potential $\bm A = M\bm\Omega\times \bm r$ in Eq.~(\ref{eqn:rot_hamilt}) 
yields an effective magnetic field $\bm B_{\mathrm{eff}}= \bm
\nabla\times \bm A = 2M \bm\Omega$ that is uniform and the
corresponding  vector potential is analogous to what is called
symmetric gauge.  (The crucial requirement for a nonzero magnetic
field is that  the vector direction of  $\bm A$ differs from that of
the intrinsic spatial dependence: for symmetric gauge, $\bm A$ is
along $\hat{\bm \phi}$ and its magnitude  depends on the distance
$r$.)   Other choices yield the same magnetic field if they are
related through a  gauge transformation $\bm A\to \bm A' = \bm A
+\bm \nabla \Lambda$, where $\Lambda$ is a scalar gauge function.
Under such a gauge transformation, the  quantum-mechanical state
vector undergoes a phase change $\psi \to \psi' =
\exp(i\Lambda/\hbar)\psi$~\cite[p.~200]{Grif05}.    For example,
another common choice takes $\bm A' = -2M \Omega\, y \,\hat{\bm x}$
(Landau gauge) where the constant coefficient is chosen so that the
two vector potentials yield the same effective uniform magnetic
field $\bm \nabla\times \bm A'$ = $2M\bm \Omega$.

The line integral of $\bm A$ plays an essential  role;  the special choice of a closed contour $\cal C$ gives the familiar gauge-invariant result $\oint_{\cal C} \bm A\cdot d\bm l = \int d\bm S\cdot \bm \nabla~\times~\bm A$, which is just the magnetic flux enclosed by the contour.  In the present case of unit fictitious charge, the flux quantum is $h$, and the enclosed flux is simply $h\times N_v $, where $N_v$ is the number of flux quanta enclosed by $\cal C$.  Equivalently, the enclosed flux is $\hbar$ times the net phase change around the closed contour (namely $\hbar \times2\pi N_v$).
In particular, if a quantum-mechanical state   yields a net phase change when it executes a closed path $\cal C$, it  experiences a gauge field $\bm A$ (either from real magnetism or from an  artificial/synthetic gauge field).

This observation leads to  the important concept of geometric phase,
often called Berry's phase~\cite[Sec.~10.2]{Berr84,Grif05}. If a
nondegenerate quantum eigenstate $|\chi_1(\bm r)\rangle$ depends on
$\bm r$ along  a contour $\cal C$, then there is a net geometric
phase change on once encircling the contour $\cal C$
 \begin{equation}\label{Berry}
\gamma_{\cal C} = \hbar^{-1} \oint_{\cal C} d \bm l\cdot \bm A(\bm r),
\end{equation}
where the integrand
\begin{equation}\label{Asyn}
\bm A(\bm r)  \equiv i\hbar \langle \chi_1(\bm r)|\bm\nabla\chi_1(\bm r)\rangle
\end{equation}
is a synthetic or artificial vector potential that  acts just like
any  familiar applied gauge field~\cite[Sec.~I.A]{Dali10}.  The line
integral $\gamma_{\cal C}$ is real and can be interpreted as the
Berry's phase~\cite{Berr84} acquired by an atom after   adiabatic
transport around  the closed loop $\cal C$ in coordinate space. This
quantity is called {\it geometric} phase because it depends only on
the path and not on the speed (assumed slow).

To make a connection with cold atoms, it is helpful to summarize a simple toy model used in~\cite[Sec.~I]{Dali10}.  Specifically, consider a two-state atom with  bare  states $|g\rangle$ and  $|e\rangle$, and a one-body Hamiltonian
$p^2/2M + V_{\rm tr}$ that is diagonal in this basis. Add  a coupling operator $U$ that reflects, for example, electromagnetic coupling between the two bare states.  Assume the general hermitian traceless form
\begin{equation}\label{U}
U = \frac{\hbar\Omega}{2}\left(\begin{matrix} \cos\theta&e^{-i\phi}\sin\theta\\
        e^{i\phi}\sin\theta& -\cos\theta
        \end{matrix}\right),
\end{equation}
where $\Omega$ is a generalized Rabi frequency [compare Eq.~(\ref{coupled})], $\theta$ is a mixing angle and $\phi$ determines the phase of the off-diagonal coupling terms. The $2\times 2$ matrix is the product $\hat{\bm n}\cdot\bm \sigma$ where the unit vector $\hat{\bm n}$ is characterized by the spherical polar angles $\theta,\phi$, and $\bm \sigma= \sum_{j=1}^3\hat{\bm x}_j\sigma_j$ represents the  Pauli matrices.   All three parameters $\Omega,\ \theta$, and $\phi$ can depend on  position $\bm r$.

The operator $U$ has two eigenvectors
\begin{equation}\label{eigen}
|\chi_1\rangle = \left(\begin{matrix} \cos(\theta/2)\\ e^{i\phi}\sin(\theta/2)\end{matrix}\right) \quad
|\chi_2\rangle = \left(\begin{matrix} -e^{-i\phi}\sin(\theta/2)\\ \cos(\theta/2)\end{matrix}\right)
\end{equation}
with eigenvalues $\hbar\Omega/2$ and $-\hbar\Omega/2$, respectively; these eigenvectors are usually known as dressed states and are the eigenstates of the atoms in the radiation field.  They can serve as the basis for an expansion of the full state vector of the particle:
\begin{equation}\label{Hilbert}
|\Psi(\bm r,t)\rangle = \sum_{j=1}^2\psi_j(\bm r,t)|\chi_j(\bm r)\rangle.
\end{equation}
Consider the action of the momentum operator $\bm p= -i\hbar \bm \nabla$ on the  state $|\Psi(\bm r,t)\rangle$.  Since the state vector (\ref{Hilbert}) is a product of two position-dependent functions, this operation produces  $\bm p|\Psi\rangle = \sum_{j=1}^2 \left[(\bm p \psi_j)|\chi_j\rangle +\psi_j\bm p|\chi_j\rangle\right]$. A little manipulation yields
 \begin{equation}\label{p-A}
\bm p|\Psi\rangle = \sum_{j,k=1}^2\left(\bm p\delta_{jk}-\bm A_{jk} \right)\psi_k| \chi_j\rangle
\end{equation}
with $\bm A_{jk} = i\hbar \langle \chi_j|\bm \nabla \chi_k\rangle $ a $2\times 2$  vector matrix in the dressed-state indices, and we have inserted the $2\times 2$ completeness relation to obtain the second  term in Eq.~(\ref{p-A}).

If the system is prepared with the particle in state $|\chi_1\rangle$, and the particle moves sufficiently slowly, it will adiabatically remain in this state.
The projection of the kinetic-energy term   $(p^2/2M)|\Psi\rangle $  onto the state $|\chi_1\rangle$ yields the effective Schr\"{o}dinger equation for the amplitude $\psi_1$ describing the center-of-mass motion in the first internal state
\begin{equation}\label{psi1}
i\hbar\frac{\partial\psi_1}{\partial t} = \left[ \frac{\left(\bm p - \bm A\right)^2}{2M} + V_{\rm tr} +\frac{\hbar\Omega}{2} + W\right]\psi_1,
\end{equation}
where the induced vector potential $\bm A$ is given in Eq.~(\ref{Asyn}) and $W= |A_{12}|^2/2M$ is an induced scalar potential arising from the elimination of the second atomic state $|\chi_2\rangle$.  Note that this synthetic gauge field $\bm A$ means that cold  neutral atoms can serve to  study   magnetic effects usually associated with electromagnetic charge.

 In the present case, the synthetic vector potential has the form $\bm A(\bm r )= \frac{1}{2}\hbar (\cos\theta-1)\bm\nabla \phi$, and the effective magnetic field becomes
$\bm B(\bm r) = \bm\nabla \times \bm A =  \frac{1}{2}\hbar \bm\nabla(\cos\theta)\times \bm\nabla \phi$.  For a  nonzero effective magnetic field,   $\bm \nabla \theta$ and $\bm \nabla \phi$ must not be collinear.

Among the many proposals for experimental implementation of these ideas~\cite[Sec.~II]{Dali10}, we here focus on  one scheme~\cite{Spie09} that has proved practical in creating vortices with a synthetic gauge potential and no externally applied rotation~\cite{Lin09}.  Two slightly detuned laser beams counter-propagate along $\hat{\bm x}$ with a difference wave vector $ q \hat{\bm x}$.  Together these beams induce Raman transitions, and the corresponding Rabi frequency has a spatial dependence $\propto e^{iqx}$ that appears as the phase $\phi=  qx$  of the coupling matrix $U$ in Eq.~(\ref{U}).  In addition, a uniform magnetic Zeeman field along $\hat{\bm y}$ provides detuning away from the resonant coupling and determines the mixing angle $\theta$.   Detailed analysis in momentum space for this two-level system~\cite{Spie09}   shows that  the minimum in the dispersion relation shifts away from $k_x=0$ to a finite value with an approximate dispersion relation $(\hbar k_x- A_x)^2/2M$. (Reference~\cite[Sec.~II.D]{Dali10} gives the corresponding real-space analysis.)  This shift yields  what is effectively a synthetic vector potential $A_x$ that depends on $q$  and on the Zeeman field.  $A_x$  would be  uniform in the absence of additional perturbations.   In the presence of  a time-dependent  $ A_x(t)$, however, the derivative $-\partial A_x/\partial t$ induces an effective electric field $E_x$.   Reference~\cite{Lin10} ramps a uniform $\bm A$ from an initial value to a final value and measures the resulting pulsed electric field, obtaining values for both the canonical momentum $\bm p$ and the mechanical momentum $M\bm v = \bm p-\bm A$.

\begin{figure}[ht]
\includegraphics[width=3in]{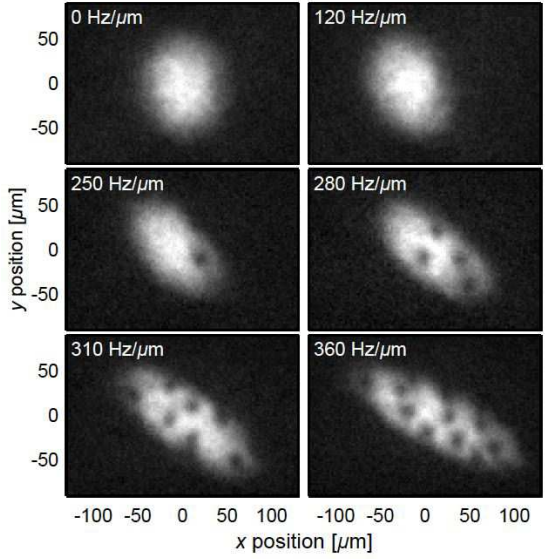}
  \caption{\label{Lin}  Vortex formation for increasing applied magnetic field  gradient 
   with $N=1.4\times 10^5$ atoms.  The six images  show the $|m_F \rangle =0$ component, and increasing magnetic field gradient is equivalent to increasing synthetic gauge potential.   Reprinted with permission of I. B. Spielman.}
  \end{figure}

 Application of a variable  magnetic field gradient   along $\hat{\bm y}$   supplies the necessary spatial variation $A_x(y)$ like the
 Landau gauge, and the combination of $\phi(x)$ and $\theta(y)$ together  produce an effectively uniform rotation (synthetic magnetic field) over a reasonably broad spatial range. In practice the actual experiment involves all three $|m_F\rangle = 1,0, -1$ magnetic substates, and the theory requires numerical analysis.  Nevertheless, Ref.~\cite{Lin09} was able to produce vortices without rotation, as shown in Fig.~\ref{Lin}. One puzzle is that the vortices do not form a regular array, in contrast to general theoretical predictions, but this may arise from insufficient time to reach equilibrium.

This field is developing rapidly, especially in connection with trapped atoms in optical lattices (see Chaps.\ 1 and 5 for general discussions).  Reference~\cite{Dali10} reviews various proposals for such  synthetic gauge fields.   Recently, Bloch's group in Munich has studied  a two-dimensional $^{87}$Rb BEC in a square lattice, producing strong effective magnetic fields with staggered (alternating striped) flux of order 1/2 flux quantum per plaquette~\cite{Aide11}.  This achievement is a significant step toward realizing  highly correlated states that are analogs of fractional quantum-Hall states~\cite{cooper2008}.

The situation becomes much more interesting if two or more degenerate states contribute simultaneously to the Berry's phase~\cite[Sec.~III]{Dali10}.  Instead of the previous example of two well-separated states $|g\rangle$ and $|e\rangle$, consider a set of $q$ degenerate ground states $|\chi_j\rangle$.  If the system starts in a linear combination of these states, then the adiabatic motion of the particle will remain within this degenerate set, leading to the following matrix  generalization of the synthetic gauge field in Eq.~(\ref{Asyn}):
\begin{equation}\label{Anon}
\bm A_{jk} = i\hbar\langle \chi_j(\bm r)|\bm \nabla \chi_k(\bm r)\rangle,
\end{equation}
where the effective vector potential is now a $q\times q$ matrix in the internal space, as well as a vector in three-dimensional coordinate space.  In general, these matrices do not commute, which leads to non-Abelian gauge fields~\cite{Wilc84} that are qualitatively different from the Abelian case with $q=1$.

One unusual aspect of these non-Abelian matrix gauge fields appears in the effective magnetic field $\bm B$. By definition, the velocity  is the combination $\bm v = (\bm p-\bm A)/M$.  Its Heisenberg equation of motion $i\hbar \dot{\bm v} = \left[\bm v,\frac{1}{2}Mv^2\right]$ leads to the expected  form $M\dot{\bm v} =\frac{1}{2}\left( \bm v\times \bm B-\bm B\times\bm v\right)$, where the symmetrized structure arises because  $\bm v$ and $\bm  B$ need not commute.  Here, the effective magnetic field  $\bm B$ is   given by the generalized curvature
\begin{equation}\label{curve}
B_i =\frac{1}{2} \epsilon_{ijk} F_{jk},\quad \hbox{where} \quad F_{jk} = \partial_jA_k - \partial_kA_j -\frac{i}{\hbar}[A_j,A_k].
\end{equation}
Even if these  gauge fields are spatial constants, the effective magnetic field does not, in general, vanish because of the term $\frac{1}{2}\epsilon_{ijk}[A_j,A_k] = (\bm A\times \bm A)_i$.  This situation is dramatically different from that of the more familiar Abelian gauge field, where the magnetic field vanishes for a spatially constant $\bm A$.  Reference~\cite{Ho10} analyzes a particular generalization of the experiments by the Spielman group involving two degenerate ground states obtained by balancing the linear and quadratic Zeeman effect along with the applied field gradient.

Another remarkable feature of these non-Abelian gauge fields is the possibility of generating an effective spin-orbit coupling for trapped neutral atoms~\cite{Ho10,Wang10,Spie10}.    In certain situations, the matrix vector potential $\bm A$ can lie in the $xy$ plane and reduce to a linear combination of the $2\times 2$ Pauli matrices $\sigma_x$ and $\sigma_y$.  Depending on the vector coefficients, the term $(\bm p\cdot \bm A + \bm A\cdot\bm p)/2M$  in the Hamiltonian can produce a spin-orbit coupling involving products of momentum components  and Pauli matrices.  Similar coupling terms $\bm p\times \bm \sigma \cdot\hat{\bm z} = p_x\sigma_y-p_y\sigma_x$ and  $p_y\sigma_y-p_x\sigma_x$  are familiar in the  condensed-matter fields of semiconductor spintronics and topological insulators, where they are known as Rashba and Dresselhaus spin-orbit coupling~\cite{Schl06}.  In this way, it  may be possible to use cold atoms to explore some of the properties of these rather complicated semiconductors, although current experiments include only a single Pauli spin matrix and thus provide only an Abelian coupling~\cite{Spie10}.  There are also proposals for $3\times 3$ spin-1 generalizations of the Pauli matrices for spin-$\frac{1}{2}$.
Reference~\cite[Sec.~III.D and Sec.~V]{Dali10} summarizes  this fascinating field and its connection with the rapidly emerging field of spintronics.

 \section*{Acknowledgments}
A.\ F.\ thanks  G.\ Juzeli${\bar{\rm u}}$nas and I.\ Spielman for very helpful discussions about synthetic gauge fields.  Part of this work was completed at the Kavli Institute for Theoretical Physics, University of  California Santa Barbara (National Science Foundation  Grant No.\  PHY05-51164),  and A.\ F.\ is  grateful for the warm hospitality.  C.\ F.\ thanks E.\ Nugent and R.\ Williams for their contributions.






\begin{thebibliography}{99}


\bibitem{Till90} D.\ R.\ Tilley and J.\ Tilley, {\em Superfluidity and Superconductivity} (Institute of Physics, Bristol, England, 1990), third edition.

\bibitem{Feyn55} R.\ P.\ Feynman, {\em Application of quantum mechanics to liquid helium}, in Progress in Low Temperature Physics, vol.\  1,  C.\ J.\ Gorter, editor (North-Holland, Amsterdam, 1955), p.~17.

 \bibitem{Donn91}  R.\ J.\ Donnelly, {\em Quantized vortices in He II} (Cambridge University Press, Cambridge, 1991).

\bibitem{Land41} L.\ D.\ Landau, {\em The theory of superfluidity of helium II},  J.\ Phys.\ (U.S.S.R.) {\bf 5}, 71 (1941).

\bibitem{Land47} L.\ D.\ Landau, {\em On the theory of superfluidity of helium II},  J.\ Phys.\ (U.S.S.R.) {\bf 11}, 91 (1947).

\bibitem{Lifs80} E.\ M.\ Lifshitz and L.\ P.\ Pitaevskii, {\em Statistical Physics} (Pergamon Press, Oxford, 1980), part 2, chap.\ III.

\bibitem{Dalf99}  F.\ Dalfovo, S.\ Giorgini, L.\ P.\ Pitaevskii, and S.\ Stringari, {\em Theory of Bose-Einstein condensation in the alkali gases}, Rev.\ Mod.\ Phys.\ {\bf 71}, 463 (1999).

\bibitem{Pita03} L.\ Pitaevskii and S.\ Stringari, {\em Bose-Einstein Condensation} (Oxford Science Publications, Oxford, 2003).

\bibitem{Peth08} C.\ J.\ Pethick and H.\ Smith, {\em Bose-Einstein Condensation in Dilute Gases} (Cambridge University Press, Cambridge, 2008),  second edition.

\bibitem{Ande95}  M.\ H.\ Anderson, J.\ R.\ Ensher, M.\ R.\ Matthews, C.\ E.\ Wieman, and E.\ A.\ Cornell, {\em Observation of Bose-Einstein condensation in a dilute atomic vapor}, Science {\bf 269}, 198 (1995).

\bibitem{Gros61} E.\ P.\ Gross, {\em Structure of a quantized vortex in boson systems}, Nuovo Cimento {\bf 20}, 454 (1961).

\bibitem{Pita61} L.\ P.\ Pitaevskii, {\em Vortex lines in an imperfect Bose gas}, Zh.\ Eksp.\  Teor.\ Fiz.\ {\bf 40}, 646 (1961) [Sov.\ Phys.---JETP {\bf 13}, 451 (1961)].

\bibitem{Baym96} G.\ Baym and C.\ J.\ Pethick, {\em Ground-state properties of magnetically trapped Bose-condensed rubidium gas}, Phys.\ Rev.\ Lett.\ {\bf 76}, 6 (1996).

\bibitem{Bogo47}  N.\ N.\ Bogoliubov, {\em On the theory of superfluidity}, J.\ Phys.\ (U.S.S.R.) {\bf 11}, 23 (1947).

\bibitem{Ozer05}  R.\ Ozeri, N.\ Katz, J.\ Steinhauer, and N.\ Davidson, {\em Bulk Bogoliubov excitations in a Bose-Einstein condensate}, Rev.\  Mod.\ Phys.\ {\bf 77}, 187 (2005).

\bibitem{Rupr95}  P.~A.~Ruprecht, M.~J.~Holland, K.~Burnett, and M.~Edwards, {\em Time-dependent solution of the nonlinear Schr\"odinger equation for Bose-condensed trapped neutral atoms},  Phys.~Rev.~A {\bf 51}, 4704 (1995).

\bibitem{Sack99}  C.\ A.\ Sackett, J.\  M.\  Gerton, M.\ Welling, and R.\ G.\ Hulet, {\em Measurements of collective collapse in a Bose-Einstein condensate with attractive interactions}, Phys.\ Rev.\ Lett.\  {\bf 82}, 876 (1999).


\bibitem{Soko95}  P.\ E.\ Sokol, {\em Bose-Einstein condensation in liquid helium}, in  {\em Bose-Einstein Condensation} (Cambridge University Press, Cambridge, 1995), ed. A.\ Griffin, D.\ W.\ Snoke, and S.\ Stringari, Chap.~4.

\bibitem{Ensh96}  J.\ R.\ Ensher, D.\ S.\ Jin, M.\ R.\ Matthews, C.\ E.\ Wieman, and E.\ A.\ Cornell, {\em Bose-Einstein condensation in a dilute gas:  measurement of energy and ground-state occupation}, Phys.\ Rev.\ Lett.\ {\bf 77}, 4984 (1996).

\bibitem{Tammuz2011}  N.\ Tammuz, R.\ P.\ Smith, R.\ Campbell, S.\ Beattie, S.\ Moulder, J.\ Dalibard, and Zoran Hadzibabic. {\em Can a Bose Gas Be Saturated?} Phys.\ Rev.\ Lett.\ {\bf 106}, 230401 (2011).


\bibitem{Hechenblaikner2000} G.\ Hechenblaikner, O.\  Marag{\`o}, E.\ Hodby, J.\ Arlt, S.\ Hopkins, and C.\ Foot,  {\em Observation of harmonic generation and nonlinear coupling in the collective dynamics of a Bose-Einstein condensate},  Phys.\ Rev.\ Lett.\ {\bf 85}, 692 (2000).

\bibitem{Nugent2003} E.\ Nugent, D.\ McPeake, and J.\ McCann, {\em Superfluid toroidal currents in atomic condensates},  Phys.\ Rev.\ A {\bf 68}, 063606 (2003).

\bibitem{Pita97} L.\ P.\ Pitaevskii, and A. Rosch, {\em Breathing modes and hidden symmetry of trapped atoms in two dimensions}, Phys.\  Rev.\ A {\bf  55}, R853 (1997).

\bibitem{Laha09}  T.\ Lahaye, C.\ Menotti, L.\ Santos, M\ Lewenstein, and T.\ Pfau, {\em The physics of dipolar bosonic quantum gases}, Rep.\ Prog.\ Phys.\ {\bf 72}, 126401 (2009).

\bibitem{Grie05}  A.\ Griesmaier, J.\ Werner, S.\ Hensler, J.\ Stuhler, and T.\ Pfau, {\em Bose-Einstein condensation of chromium}, Phys.\ Rev.\ Lett.\ {\bf 94}, 160401 (2005).

\bibitem{Lu11} M\. Lu, N.\ Q.\ Burdick, S.\ H.\ Youn, and B.\ L.\ Lev, {\em Strongly dipolar Bose-Einstein condensate of dysprosium}, Phys.\ Rev.\ Lett.\ {\bf 107}, 190401 (2011).

\bibitem{Ospe08}  S.\ Ospelkaus, A.\ Pe'er, K.-K.\ Ni, J.\ J.\ Zirbel, B.\  Neyenhuis, S.\ Kotochigova, P.\ S.\ Julienne, J.\ Ye, and D.\ S.\ Jin, {\em Efficient state transfer in an ultracold dense gas of heteronuclear molecules}, Nature Phys.\  {\bf 4}, 622 (2008).

\bibitem{Ni08}   K.-K.\ Ni, S.\ Ospelkaus, M.\ G.\  H.\ de Miranda, A.\ Pe'er, B.\  Neyenhuis,  J.\ J.\ Zirbel, S.\ Kotochigova, P.\ S.\ Julienne,  D.\ S.\ Jin, and J.\ Ye, {\em A high phase-space-density gas of polar molecules}, Science {\bf 322}, 231 (2008).

\bibitem{Carr09}  L.\ D.\ Carr, D.\ DeMille, R.\ V.\ Krems, and J.\ Ye, {\em Cold and ultracold molecules:  science, technology and applications}, New J.\ Phys.\ {\bf 11}, 055049 (2009).

\bibitem{Koch08}  T.\ Koch, T.\ Lahaye, J.\ Metz, B.\ Fr{\"o}hlich, A.\ Griesmaier, and T.\ Pfau, {\em Stabilization of a purely dipolar quantum gas against collapse}, Nature Phys.\  {\bf 4}, 218 (2008).

\bibitem{Laha08}   T.\ Lahaye, J.\ Metz, B.\ Fr{\"o}hlich, T.\ Koch, M.\  Meister, A.\ Griesmaier, T.\ Pfau, H.\ Saito, Y.\ Kawaguchi, and M.\ Ueda, {\em $d$-Wave collapse and explosion of a dipolar Bose-Einstein condensate}, Phys.\ Rev.\ Lett.\ {\bf 101}, 080401 (2008).

\bibitem{Matt98}  M.\ R.\ Matthews, D.\ S.\ Hall, D.\ S.\ Jin, J.\ R.\ Ensher, C.\ E.\ Wieman, E.\ A.\ Cornell, F.\ Dalfovo, C.\ Minniti, and S.\ Stringari, {\em Dynamical response of a Bose-Einstein condensate to a discontinuous change in internal state}, Phys.\ Rev.\ Lett.\ {\bf 81}, 243 (1998).

\bibitem{Feyn65}  R.\ P.\ Feynman, R.\ B.\ Leighton, and M.\ Sands, {\em The Feynman Lectures on Physics  Quantum Mechanics} (Addison-Wesley, Reading, MA, 1965), Vol.\ III.

\bibitem{Grif05}  D.\ J.\ Griffiths, {\em Introduction to Quantum Mechanics} (Pearson Prentice Hall, Upper Saddle River, NJ 2005), second edition.

\bibitem{Hall98}  D.\ S.\ Hall, M.\ R.\ Matthews, C.\ E.\ Wieman, and E.\ A.\ Cornell, {\em Measurement of relative phase in two-component Bose-Einstein condensates}, Phys.\ Rev.\ Lett.\ {\bf 81}, 1543 (1998).

\bibitem{Matt99}   M.\ R.\ Matthews, B.\ P.\ Anderson, P.\ C.\ Haljan, D.\ S.\ Hall,  C.\ E.\ Wieman, and  E.\ A.\ Cornell {\em Vortices in a Bose-Einstein condensate}, Phys.\ Rev.\ Lett.\ {\bf 83},  2498 (1999).

\bibitem{Stam98}  D.\ M.\ Stamper-Kurn, M.\ R.\ Andrews, A.\ P.\ Chikkatur, S.\ Inouye, H.-J.\ Miesner, J.\ Stenger,  and W.\ Ketterle, {\em Optical confinement of a Bose-Einstein condensate}, Phys.\ Rev.\ Lett.\ {\bf 80},  2027 (1998).

\bibitem{Ho98} T.-L.\ Ho, {\em Spinor Bose condensates in optical traps}, Phys.\ Rev.\ Lett.\ {\bf 81},  742 (1998).

\bibitem{Ohmi98}  T.\ Ohmi and K.\  Machida, {\em Bose-Einstein condensation with internal degrees of freedom in alkali atom gases}, J.\ Phys.\  Soc.\ Jpn.\   {\bf 67}, 1822 (1998).

\bibitem{Sten98}  J.\ Stenger,  S.\ Inouye, D.\ M.\ Stamper-Kurn,  H.-J.\ Miesner,  A.\ P.\ Chikkatur,   and W.\ Ketterle, {\em Spin domains in ground-state   Bose-Einstein condensates}, Nature {\bf 396}, 345 (1998).

\bibitem{Stam00}  D.\ M.\  Stamper-Kurn and W.\ Ketterle, {\em Spinor condensates and light scattering from Bose-Einstein condensates},  in {\em Coherent Atomic Matter Waves},  Proceedings of the Les Houches Summer School, Session LXXII, 1999, edited by R.\ Kaiser, C.\ Westbrook, and F.\ David (Springer, New York, 2001), p.\ 137.  This review is available as
   e-print: {\it cond-mat} 0005001.

\bibitem{Veng08}  M.\ Vengalattore, S.\ R.\ Leslie, J.\ Guzman, and D.\ M.\ Stamper-Kurn, {\em Spontaneously modulated spin textures in a dipolar spinor Bose-Einstein condensate},  Phys.\ Rev.\ Lett.\ {\bf 100},  170403 (2008).

\bibitem{Guery-Odelin1999} D.\ Guery-Odelin, and S.\ Stringari  {\em Scissors mode and superfluidity of a trapped Bose-Einstein condensed gas},  Phys.\ Rev.\ Lett.\ {\bf 83}, 4452 (1999).

\bibitem{Marago2000} O.\  Marag{\`o},  S.\ Hopkins, J.\ Arlt, E.\ Hodby, G.\ Hechenblaikner, and C.\ Foot, {\em Observation of the scissors mode and evidence for superfluidity of a trapped Bose-Einstein condensed gas},  Phys.\ Rev.\ Lett.\ {\bf 84}, 2056 (2000).


\bibitem{Anderson2000} B.\ P.\ Anderson, P.\ C.\ Haljan, C.\ E.\ Wieman, and E.\ A.\ Cornell,  {\em Vortex Precession in Bose-Einstein Condensates: Observations with Filled and Empty Cores},  Phys.\ Rev.\ Lett.\ {\bf 85}, 2857 (2000).

\bibitem{Codd04}  I.\ Coddington, P.\ C.\ Haljan, P.\  Engels, V.\  Schweikhard, S.\ Tung, and E.\ A.\ Cornell, {\em Experimental studies of equilibrium vortex properties in a Bose-condensed gas}, Phys.\ Rev.\  A {\bf 70}, 063607 (2004).

\bibitem{williams2010} R.\ A.\ Williams, S.\ Al-Assam, and C.\ J.\ Foot, {\em Observation of Vortex Nucleation in a Rotating Two-Dimensional Lattice of Bose-Einstein Condensates}, Phys.\ Rev.\ Lett.\ \textbf{104}, 050404 (2010).


  
\bibitem{Zwierlein05}  M.\ W.\ Zwierlein, J.\ R.\ Abo-Shaeer, A.\ Schirotzek, C.\ H.\ Schunck, and W.\ Ketterle. {\em Vortices and superfluidity in a strongly interacting Fermi gas} Nature {\bf 435}, 1047 (2005).



\bibitem{Scherer07}  D.\ R.\ Scherer, C.\ N.\ Weiler, T.\ W.\ Neely, and B.\ P.\ Anderson. {\em Vortex formation by merging of multiple trapped Bose-Einstein condensates} Phys.\ Rev.\ Lett.\ {\bf 98}, 110402 (2007).

\bibitem{Dali10} J.\  Dalibard, F.\  Gerbier, G.\ Juzeli$\rm{\bar u}$nas, and P.\  \"{O}hberg,   {\em Artificial gauge potentials for neutral atoms}, Rev.\ Mod.\ Phys.\ {\bf 83}, 1523 (2011).

\bibitem{cooper2008} N.\ R.\ Cooper, {\em Rapidly Rotating Atomic Gases}, Advances in Phys.\ \textbf{57}, 539 (2008).

\bibitem{schweikhard2004} V.\ Schweikhard, I.\ Coddington, P.\ Engels, V.\ Mogendorff, and E.\ A.\ Cornell,  {\em Rapidly Rotating Bose-Einstein Condensates in and near the Lowest Landau Level}, Phys.\ Rev.\ Lett.\  \textbf{92}, 040404 (2004).



\bibitem{Berr84}  M.\ V.\ Berry, {\em Quantal phase factors accompanying adiabatic changes}, Proc.\ R.\ Soc.\ Lond.\ A {\bf 392}, 45  (1984).

\bibitem{Spie09}  I.\ B.\ Spielman, {\em Raman processes and effective gauge potentials}, Phys.\  Rev.\ A {\bf 79}, 063613 (2009).

\bibitem{Lin09}  Y.-J.\ Lin, R.\ L.\ Compton, K.\ Jim\'{e}nez-Garc\'{\i}a, J.\  V.\ Porto, and I.\ B.\  Spielman, {\em Synthetic magnetic fields for ultracold neutral atoms},  Nature {\bf 462}, 628 (2009).

\bibitem{Lin10}  Y.-J.\ Lin, R.\ L.\ Compton, K.\ Jim\'{e}nez-Garc\'{\i}a, W.\ D.\ Phillips, J.\  V.\ Porto, and I.\ B.\  Spielman, {\em A synthetic electric force acting on neutral atoms},    Nature Physics {\bf 7}, 531 (2011), doi:10.1038/nphys1954.

\bibitem{Aide11}  M.\ Aidelsburger, M.\ Atala, S.\ Nascimb{\`e}ne, S.\ Trotzky, Y.-A.\ Chen, and I.\ Bloch, {\em Experimental realization of strong effective magnetic fields in an optical lattice}, Phys.\ Rev.\ Lett.\  {\bf 107}, 255301 (2011).

\bibitem{Wilc84} F. Wilczek and A. Zee, {\em Appearance of gauge structures in simple dynamical systems}, Phys.\ Rev.\ Lett.\ {\bf 52},  2111 (1984).

\bibitem{Ho10}  T.-L.\ Ho and S.\ Zhang, {\em Bose-Einstein condensates with spin-orbit interaction}, Phys.\ Rev.\ Lett.\ {\bf 107},  150403 (2011).

\bibitem{Wang10}  C.\  Wang, C.\ Gao, C.-M.\ Jian, and H.\ Zhai, {\em Spin-orbit coupled spinor Bose-Einstein condensates},  Phys.\ Rev.\ Lett.\ {\bf 105},  160403 (2010).

\bibitem{Spie10} Y.-J.\ Lin, K.\  Jim{\'e}nez-Garc{\'\i}a, and  I.\ B.\  Spielman, {\em Spin-orbit-coupled Bose-Einstein condensates}, Nature {\bf 471}, 83 (2011).

\bibitem{Schl06}  J.\ Schliemann, D.\ Loss, and R.\ M.\ Westervelt, {\em {\rm Zitterbewegung} of electrons and holes in III-V semiconductor quantum wells},  Phys.\ Rev.\ B {\bf 73}, 085323 (2006).



 \end{thebibliography}



\end{document}